\documentclass[prodmode,acmtissec]{acmsmall} 

\usepackage{amssymb}
\usepackage{graphicx}
\usepackage{amsmath}
\usepackage{psfrag}
\usepackage{multirow}
\usepackage{url}
\usepackage[usenames,dvipsnames,svgnames,table]{xcolor}

\usepackage[ruled]{algorithm2e}

\SetAlFnt{\small}
\SetAlCapFnt{\small}
\SetAlCapNameFnt{\small}
\SetAlCapHSkip{0pt}
\IncMargin{-\parindent}

\acmVolume{0}
\acmNumber{0}
\acmArticle{0}
\acmYear{2016}
\acmMonth{6}

\setcopyright{none}

\begin{document}

\markboth{L. Mu\~noz-Gonz\'{a}lez et al.}{Efficient Attack Graph Analysis through Approximate Inference}

\title{Efficient Attack Graph Analysis through Approximate Inference}
\author{LUIS MU\~NOZ-GONZ\'{A}LEZ
\affil{Imperial College London}
DANIELE SGANDURRA
\affil{Imperial College London}
ANDREA PAUDICE
\affil{Imperial College London}
EMIL C. LUPU
\affil{Imperial College London}}

\begin{abstract}
Attack graphs provide compact representations of the attack paths that an attacker can follow to compromise network resources by analysing network vulnerabilities and topology. These representations are a powerful tool for security risk assessment. Bayesian inference on attack graphs enables the estimation of the risk of compromise to the system's components given their vulnerabilities and interconnections, and accounts for multi-step attacks spreading through the system. Whilst static analysis considers the risk posture at rest, dynamic analysis also accounts for evidence of compromise, e.g. from SIEM software or forensic investigation. However, in this context, exact Bayesian inference techniques do not scale well. In this paper we show how Loopy Belief Propagation - an approximate inference technique - can be applied to attack graphs, and that it scales linearly in the number of nodes for both static and dynamic analysis, making such analyses viable for larger networks. We experiment with different topologies and network clustering on synthetic Bayesian attack graphs with thousands of nodes to show that the algorithm's accuracy is acceptable and converge to a stable solution. We compare sequential and parallel versions of Loopy Belief Propagation with exact inference techniques for both static and dynamic analysis, showing the advantages of approximate inference techniques to scale to larger attack graphs.
\end{abstract}

%
%
%

%
%

\terms{Security risk assessment, attack graphs, dynamic analysis}



\begin{bottomstuff}
This work has been supported by the UK government under EPSRC grant EP/L022729/1.

Authors' address: Department of Computing, Imperial College London, 180 Queen's Gate, SW7 2AZ, London, UK. E-mail: \{l.munoz, d.sgandurra, a.paudice15, e.c.lupu\}@imperial.ac.uk.
\end{bottomstuff}

\maketitle

\section{Introduction}
Despite significant efforts to protect networks against cyber-attacks \cite{gartner}, system administrators cannot cope with the sophistication of modern threats, as shown by the history of breaches that organizations have suffered in recent years \cite{digitalguardian}. One of the most common strategies to protect networks is to identify and patch vulnerabilities. However, this is often not systematically done, either for lack of manpower or because it requires interrupting critical systems. A risk-driven approach is therefore needed to optimise resources for network protection. Such an approach requires assessing the networks risks, prioritizing the most critical threats, and then estimating the \emph{risk exposures}, given the likelihood of threats and the severity of the impacts \cite{wheeler}. Finally, these values are used by administrators to select appropriate countermeasures.
But often this analysis is carried out separately for each of the network components ignoring interdependencies between vulnerabilities, i.e. how successfully exploiting a vulnerability allows an attacker to exploit other vulnerabilities, thus moving across the network and acquiring privileges at every step. 

These shortcomings can be addressed using Attack Graphs (AGs) \cite{sheyner2002,albanese}, a well-established technique to represent the possible paths of an attacker through the system by exploiting successive vulnerabilities. AGs allow system administrators to reason about threats and risks in a formal way to better select countermeasures \cite{ingols}. Two types of analysis can be undertaken. \textit{Static analysis} determines the \textit{a priori} risks to which network components are exposed. \textit{Dynamic analysis} updates those risks in light of any indication that some of the network components may have been compromised, e.g. from Security Information and Event Management (SIEM) and Intrusion Detection Systems (IDS). Dynamic analysis also allows administrators to profile the attacker's paths, to determine the nodes that are more likely to be attacked in the next steps. This enables administrators to evaluate the security risk for valuable resources in the network and reason about the nodes that may have been already compromised when we observe evidence of an ongoing attack. 
As organizations are often under attack, dynamic analysis gives administrators important insights on the most vulnerable targets and where they should spend their efforts at run-time. Although several metrics have been proposed to perform security risk assessment using AGs \cite{idika,noelMetrics}, taking into consideration the length and the number of paths that let the attacker reach a goal, or the global impact of the existing vulnerabilities in the network, they fail to consider the difficulty of exploiting each vulnerability and the dependencies between the different attack paths. In this sense, it is easy to observe that both static and dynamic analysis of AGs have inherent probabilistic characteristics given the uncertainty about the attackers' ability to successfully exploit vulnerabilities. Therefore, considering the dependencies between vulnerabilities, Bayesian Networks (BNs) provide an appropriate framework to model AGs, since they depict causal relationships between random variables in a compact way, so that they can model the uncertainty about the attacker's behaviour and capabilities.

Bayesian Attack Graphs (BAGs) can be analysed through efficient algorithms, such as Variable Elimination or Junction Tree (JT), to make \emph{exact inference}, i.e. to compute the unconditional probabilities of all the nodes in the BAG, i.e. the probabilities that an attacker can reach the different security states in the AG \cite{luis}. 
However, computing these probabilities in BNs is known to be an NP-Hard problem \cite{cooper}, and this limits the applicability of exact inference techniques to medium-size graphs (in the order of 100-1,000 nodes), especially when the structure of the graph is dense \cite{luis}. However, empirical investigations show that networks are highly complex: estimates of mean corporate network size are in the order of thousand of nodes \cite{5958263,6565238}. Moreover, each host may have somewhere between 2 and 11 vulnerabilities according to \cite{whitehatsec_report}. For example, Websites are reported to have an average of 6.5 vulnerabilities \cite{threat_report}. This impacts the size of the AGs, as the number of vulnerabilities also determines the number of potential attack paths. In this context exact inference techniques would be very slow and computationally very expensive, which limits their applicability for practical purposes, especially for the dynamic analysis of BAGs. For this reason, simpler metrics, computationally less demanding, have been proposed for tractable analysis of AGs in real networks \cite{idika,noelMetrics}. However, these metrics do not consider the dependencies between vulnerabilities.

To sidestep this limitation we propose to use \emph{approximate inference} techniques to allow us to analyse AGs in larger networks in a tractable way. Although making approximate inference in BNs is also NP-Hard \cite{koller}, approximate inference algorithms, such as Loopy Belief Propagation (LBP) \cite{pearl}, have better scalability than exact inference techniques. We should not be deterred by the ``approximation'' involved in this context for two reasons. First, because the probabilities are used for prioritising threats, so even if significant differences between them are meaningful, their absolute or accurate values are relatively less important. Second, because the probability of successful exploitation of a vulnerability is already a rough estimate, often based on Common Vulnerability Scoring System (CVSS) \cite{cvss}, which ignores other factors such as attacker's skills, knowledge and tooling and, hence, they may not be an accurate indication of compromise \emph{per se}. Therefore, notwithstanding their inherent approximation, the better scalability of these algorithms makes them highly crucial to perform qualitative risk analysis of large networks. Although the validity of the results is limited by the errors in the estimates of the probabilities, in our experimental evaluation we show that the accuracy of the probability estimates provided by LBP is reasonable for static and dynamic risk analysis and mitigation using BAGs.

The main contributions of this work are as follows:
\begin{itemize}
\item To the best of our knowledge, we are the first to propose the use of approximate inference techniques for the scalable analysis of AGs. Furthermore, we propose and compare both the sequential and parallel implementations of LBP to estimate the probabilities of compromise for the network nodes. 
\item We provide a comprehensive experimental evaluation using synthetic AGs which emulate the complexity of real scenarios. These experiments allow us to assess the applicability of sequential and parallel implementations of LBP for the analysis of AGs with thousands of nodes, and show the limitations of existing exact inference approaches, such as the JT algorithm \cite{shenoy}.
\item We show through experiments that LBP scales linearly in the number of nodes for both static and dynamic analysis of BAGs, which contrasts with the exponential scalability of exact inference (e.g. JT) in some cases, especially when the AG is dense. The experiments further show that the accuracy of LBP is sufficient for many practical needs exhibiting a rooted mean squared error smaller than $0.03$. 
\item Finally, we show that it is possible to get accurate results before the LBP algorithm fully converges, by allowing administrators to monitor the probability estimates at each iteration, and enabling them to start planing risk mitigation strategies in advance. This is not possible with exact inference techniques.
\end{itemize}

The rest of the paper is organised as follows. In Section 2 we discuss the related work, and in Section 3 we describe the two typical representations of AGs. In Section 4 we introduce the Bayesian AG model, including an example of a small typical corporate network as a use case. The use of Belief Propagation and LBP for the analysis of BAGs is described in Section 5. We discuss the Junction Tree algorithm proposed in \cite{luis} in Section 6. In Section 7 we present the experimental results for the static and dynamic analysis of synthetic AGs. Finally, in Section 8 we present the main conclusions and our plans for further work.

\section{Related Work} \label{sec:RelatedWork}
AG representations are built by analysing the interdependencies between the vulnerabilities and the security conditions identified in a network. Two representations are commonly encountered in the literature: \emph{State-based} representations \cite{jha,phillips,sheyner2002,sheyner2004,swiler} depict the whole state of the network in each node in the graph whilst \emph{logical} AGs \cite{ammann,jajodia,ou2006} are bipartite graphs representing the dependencies between vulnerabilities and security conditions. Although state-based AGs contain all the possible attack paths that can allow an attacker to reach a target security condition, they scale exponentially with the number of vulnerabilities and nodes in the network, which limits their application to very small networks. Relying on the monotonicity principle, logical AGs eliminate duplicate paths and provide a more compact representation that scales polynomially with the number of vulnerabilities \cite{ammann,jajodia}.   

AGs are a powerful tool to perform risk assessment and different metrics have been proposed in the literature. \cite{lippmann} propose to use the percentage of network assets an attacker has compromised. However, this metric is not goal-oriented, as it is the case of AGs. \cite{pamula} propose the weakest adversary metric, i.e. measure the risk according to the weakest attack path in the graph. Simpler approaches are used in \cite{phillips,ortalo,li}, where they propose to use the shortest path, the number of paths, and the average path length as metrics to measure risk. In a similar way, \cite{idika} propose the normalized mean, standard deviation, mode, and median of the path lengths as a set of metrics to assess risk in AGs. In \cite{noelMetrics} propose a metrics suite for AG that takes into account the CVSS scores of the vulnerabilities in the AGs as well as topological aspects of the graph, such as the connectivity, the number of cycles, and the depth. However, most of these metrics fail to account for the dependencies of the vulnerabilities and attack paths, as well as the difficulty to exploit the vulnerabilities. These limitations can be addressed with probabilistic models, which allow to compute the probabilities of each node in the graph to be compromised by an attacker when the network is at rest or under attack. 

Probabilistic models for AGs have been already proposed in the literature: For example, \cite{frigault,wang2008} present mechanisms to calculate the conditional probability tables, which represent the probabilities of compromising a node given all possible states of the parent nodes (or preconditions). However, no mechanism is proposed to calculate the unconditional probabilities, i.e. the probabilities of compromising the nodes in the network regardless of the state of their corresponding preconditions. A more complete Bayesian model is described in \cite{xie}, which takes into account non-perfect behaviour of the alert correlation system and the impact of zero-day vulnerabilities. However, no inference technique is proposed to calculate the unconditional probabilities and the experimental evaluation does not show the applicability of their model to networks of different sizes.

Several techniques have also been proposed in the literature for exact inference on BAGs. For example, forward-backward propagation is proposed in \cite{poolsappasit} to compute the unconditional probabilities. However, this procedure is only valid for chains \cite{rabiner,murphy} and cannot be applied to general AGs. \cite{liu} propose to use Variable Elimination (VE) \cite{dechter} for exact inference on BAGs. Although this is an efficient technique, it is highly dependent on the heuristic for the elimination ordering and no heuristic is suggested in \cite{liu}. Furthermore, none of these papers reports an experimental evaluation of the time and memory required by the techniques proposed to assess their suitability for static and dynamic analysis of AGs. More recently, the JT algorithm was proposed in \cite{luis} for exact inference in BAGs. This technique allows us to efficiently compute the exact unconditional probabilities by using a probabilistic message passing scheme on a clique tree representation of the original graph \cite{shenoy,shafer}. The experimental evaluation in \cite{luis} shows the advantages of JT over VE in terms of the time and the memory required. However, the applicability of JT to large networks is limited, especially when the AGs are dense, i.e. there are a lot of attack paths. 

\cite{6787327}, \cite{baiardi_esd} present a Monte Carlo-based method to generate AGs by simulating intelligent attackers' plans. By collecting samples in these simulations, the proposed tool returns a dataset used to compute statistics of interest for the assessment, such as the success probability of the agents or their average impact. \cite{4755419} present a Bayesian framework to express AGs that enables the computation of the probability that attacks will succeed, and the corresponding expected loss given the instantiated architectural scenario.
\cite{cuppens2002correlation} propose a framework to correlate attacks with intrusion goals, and introduce the notion of anti-correlation, which is useful to decide whether a sequence of correlated actions can lead to an intrusion. Hence, it can also be used to eliminate false positives. \cite{Tan2003} introduces two algorithms to cluster hosts based on their observed connection patterns considering that they can change over time. The experimental results obtained for real networks show that the number of clusters can be two orders of magnitude smaller than the number of hosts. Therefore, we can use these clusters to produce smaller AGs that can accurately summarise the state of the network.

\section{Attack Graphs} \label{sec:attack}
AGs are graphical models that represent the knowledge about network vulnerabilities and their interactions, showing the different paths an attacker can follow to reach a given goal by exploiting a set of vulnerabilities. Along each attack path, vulnerabilities are exploited in sequence, so that each successful exploit gives the attacker more privileges towards his goal. In the literature of AGs we can distinguish two main types of representations, namely \emph{state-based} and \emph{logical} AGs.

\begin{definition}
A \emph{state-based AG} is a tuple $\text{AG} = \{ S, \tau, S_0, S_t \}$, where $S$ is a set of states, $\tau \subseteq S \times S$ is a transition relation, $S_0 \subseteq S$ is a set of initial states, and $S_t \subseteq S$ is a set of target states \cite{sheyner2002}.
\end{definition}

State-based representations of AGs \cite{jha,phillips,sheyner2002,sheyner2004,swiler} result in directed graphs, where each node represents the state of the whole network after a successful atomic attack. This approach has two main shortcomings: First, the number of states and variables combinatorially explodes when increasing the number of nodes in the network. Second, these representations contain duplicate attack paths that differ only in the order of the attack steps, which increases the complexity of the graph. This limits the applicability of state-based representations to very small networks \cite{ammann,jajodia,ou2006}.
 
The scalability problems of state-based representations are overcome with \emph{logical AGs}, which are bipartite graphs which represent dependencies between exploits and security conditions \cite{ammann,jajodia}. More formally:

\begin{definition}
A \emph{logical AG} is a directed bipartite graph $G = (E \cup C, R_r \cup R_i)$, where the vertices $E$ and $C$ are the sets of exploits and security conditions, respectively, and the edges $R_r \subseteq C \times E$ and $R_i \subseteq E \times C$  are \textit{require} and \textit{imply} relations \cite{albanese}.
\end{definition}

These representations rely on a monotonicity principle: that the attacker never relinquishes privileges once obtained. Although not always applicable, this assumption is reasonable in most cases, as discussed in \cite{ammann}. Monotonicity allows to remove duplicated paths and to generate AGs that grow polynomially with the number of vulnerabilities and the number of connected pairs of hosts \cite{albanese}.

\section{Bayesian Attack Graphs} \label{sec:BAGs}
Some of the literature on AG analysis assumes that monotonicity induces a Directed Acyclic Graph (DAG) structure of logical AGs \cite{liu,poolsappasit,luis}. Although monotonicity helps to get rid of many cycles related to duplicate attack paths (that appear in state-based representations) some cycles still remain. However, \cite{wang2008} explain how to handle and eliminate cycles without loss of integrity. In this paper, we consider AGs with a DAG structure. Where cycles appear, we refer to \cite{wang2008} to build the corresponding conditional probability tables in the nodes affected by the cycle. The DAG structure of logical AGs, the uncertainty about the attacker's behaviour and capabilities, make BNs a suitable alternative to model AGs and perform static and dynamic analysis. In particular, BNs allow us to calculate the probability of an attacker to reach a security condition (state) in the AG.

\begin{definition}
A BN is a directed acyclic graphical model where the nodes represent random variables and the directed edges represent the dependencies between random variables. Let ${\bf X} = \{ X_1, ..., X_n \}$ be a set of (continuous or discrete) random variables. The joint probability distribution can be written as:
\begin{equation}
p({\bf X}) = \prod_{i = 1}^n p(X_i | {\bf pa}_i )
\label{eq1}
\end{equation} Then, under the BN representation, for each node $X_i$ there is a directed edge from each node in ${\bf pa}_i$, the set of parents nodes of $X_i$, pointing to $X_i$.
\end{definition}

In the context of the BAG, the nodes represent the different security states that an attacker can reach. We model the behaviour of these states as Bernoulli random variables. Hence, the probability of an attacker to compromise a node $X_i$ is $\text{Pr}(X_i = T) = p$, whereas the probability of an attacker not to compromise that node is $\text{Pr}(X_i = F) = 1 - p$, with $p \in [0,1]$\footnote{To simplify the mathematical notation we will refer to the unconditional probability of a node to be compromised as $\text{Pr}(X_i)$ instead of $\text{Pr}(X_i = T)$}. The probabilities of an attacker successfully exploiting a vulnerability, needed to compute the conditional probabilities $p(X_i | {\bf pa}_i )$ in (\ref{eq1}), are represented as parameters of the model, since these values varies slowly across time (in the order of days or weeks).

\subsection{Model Assumptions} \label{sec:modelAssumptions}
In line with much of the related work \cite{liu,frigault,poolsappasit,luis}, we make some assumptions on our model: 
\begin{itemize}
\item We consider that the probability of successfully exploiting a single vulnerability does not affect the probabilities that the attacker can successfully exploit other vulnerabilities. We also assume that these probabilities remain nearly constant in time. Although in \cite{frigault} the dynamic aspects of vulnerabilities are modelled with a dynamic BN, in practice changes to the probabilities typically occur over periods of days or weeks. Therefore, we argue that it is better to recompute the model when changes occur rather than to increase the complexity of the model to include the dynamic aspects of these probabilities. 
\item We assume that the probability of successfully exploiting a vulnerability is the same regardless of the attacker. However, these probabilities can be updated according to other models that take into account the attackers' capabilities and preferences \cite{baiardi_esd}. For example, we can use different sets of AG parameters corresponding to different attacker models identified as proposed in \cite{baiardi_esd}. We can then select, according to the attacker model, the corresponding parameters to perform the dynamic analysis of the AG.
\item We also assume that the topology of the network, host connectivity (including open ports) and the set of vulnerabilities do not change during the dynamic analysis of the BAG. This would require dynamic AG generation, which is out of the scope of this paper. However, if existing vulnerabilities are patched at run-time, our BAG representation can be easily updated by setting the probability of successful exploitation of the patched vulnerabilities to zero. On the contrary, if new vulnerabilities are discovered or new nodes are added to the network, a new AG needs to be generated.
\item We do not consider zero-day vulnerabilities, social engineering attacks, and insider threats. Although it is easy to incorporate these kinds of threats in the AG by adding one extra attack path to each security condition in the graph, as proposed in \cite{poolsappasit,xie}, the difficulty lies in estimating a reasonable probability of successful exploitation of those zero-day vulnerabilities and social engineering attacks. In fact, no mechanism for the estimation of these probabilities is proposed in the literature, except from letting the network administrator quantify their effects (in a subjective way) \cite{poolsappasit,xie}, which is however a challenging task. A more detailed description of the difficulties to estimate the impact of zero-day vulnerabilities in AGs can be found in \cite{zero_day_albanese}.
\end{itemize}

\subsection{Conditional Probability Tables}
The information available at each node $X_i$ in a BAG is the conditional probability distribution $p(X_i | {\bf pa}_i)$, the probability of a node to be compromised given the state of its parent nodes ${\bf pa}_i$. Thus, these conditional probabilities represent the probabilities of an attacker to reach a security state $X_i$ given the observations of its preconditions ${\bf pa}_i$ and the vulnerabilities ${\bf v}_i$ that can be exploited to compromise $X_i$. In this sense, we consider that the probabilities of successfully exploiting vulnerabilities are parameters of the BAG model that are used to calculate $p(X_i | {\bf pa}_i)$.

A common approach to estimate $p_{v_i}$, the probability of an attacker successfully exploiting a vulnerability $v_i$, is based upon CVSS \cite{cvss}. Although CVSS scores are intended to estimate the impact of a vulnerability rather than its average probability of being successfully exploited, CVSS scores (or some of their submetrics) are often used in the literature to estimate $p_{v_i}$ \cite{frigault,poolsappasit,luis}. In this sense, the exploitability submetric of CVSS can be considered more appropriate to estimate $p_{v_i}$, since it tries to measure the difficulty of exploiting a vulnerability.

Finally, given an estimate of the probabilities of exploitation of the vulnerabilities that allow an attacker to reach a security state $X_i$ from states ${\bf pa}_i$, we consider two possible cases to build the conditional probability tables \cite{poolsappasit,luis}: \textit{AND} and \textit{OR} conditional probability tables. In the first case, all the preconditions need to be satisfied to be able to compromise $X_i$, i.e. the attacker needs to compromise all the nodes in ${\bf pa}_i$ before being able to perform an attack to compromise $X_i$. In the case of \textit{OR} conditional probability tables, the attacker only needs to compromise one of the nodes in ${\bf pa}_i$ to attempt an attack to reach the security state $X_i$. 

Considering that the alert system is not perfect (the IDS can trigger false alarms or miss events, and that some other events maybe stealthy) and that its estimated error rate is $p_e$, \textit{AND} conditional probability tables can be calculated as:
\begin{equation}
p(X_i | {\bf pa}_i) = \begin{cases} p_e, & \exists X_j \in {\bf pa}_i | X_j = F \\ 1 - (1 - p_e) \ (1 - \prod_{j: X_j} p_{v_j}), & \text{otherwise} \end{cases}
\label{eqAND}
\end{equation} whereas for \textit{OR} conditional probability tables, using the noisy-OR formulation \cite{koller}, we have:
\begin{equation}
p(X_i | {\bf pa}_i) = \begin{cases} p_e, & \forall X_j \in {\bf pa}_i | X_j = F \\ 1 - (1 - p_e) \ \prod_{j: X_j} (1 - p_{v_j}), & \text{otherwise} \end{cases}
\label{eqOR}
\end{equation}
Estimating the error probability, $p_e$, of the alert correlation system is difficult due to the dynamic aspects of the system behaviour. In this sense, several approaches estimate this error probability based on \emph{ad hoc} methodologies and test the system under particular conditions \cite{Milenkoski}. More recently, \cite{JubaMLSR15} propose a novel approach to evaluate alert correlation systems providing statistical guarantees on the measured performances. 

Combining (\ref{eqAND}) and (\ref{eqOR}) we can extend the construction of conditional probability tables to intermediate cases, where different subsets of preconditions need to be satisfied before trying to compromise $X_i$. As proposed in \cite{xie}, the non-perfect behaviour of the alert system can also be modelled by adding an extra node $O_i$ to describe the observation from the alert system. Thus, $X_i$, which describes the actual state of the node, would be the parent of $O_i$. Then, the conditional probability table $p(O_i|X_i)$ can be built taking into account the estimation of the false alarm and detection rates of the alert system.

\subsection{Prior on the Attacker's Initial State}
In AG representations, there is usually a leaf node representing the initial state of the attacker when the attacker has not compromised any node in the network yet. Following the model in \cite{luis}, we consider that this node is not really a random variable, since it only represents that the attacker has full rights on his own machine. Under the Bayesian representation, we can consider that the Bernoulli random variable $X_0$ representing the initial state of the attacker has $\text{Pr}(X_0) = 1$. Although \cite{poolsappasit} propose to use this initial node to reflect some subjective prior knowledge of the attacker capabilities (by letting the administrator set the value of $\text{Pr}(X_0)$), this can lead to misleading conclusions, especially when reasoning using new evidence about the nodes that an attacker may have already compromised, as discussed in \cite{luis}. Finally, to obtain more accurate estimations of the unconditional probabilities needed for the analysis of BAGs, we can break loops in the BN by instantiating one initial node for each possible initial attack path. This does not affect the value of the unconditional probabilities of the rest of the nodes but can favour the convergence and the accuracy of LBP estimates \cite{murphyLBP}.

\subsection{Static and Dynamic Analysis of the BAG}
For the static analysis of AGs we are interested in calculating the \emph{unconditional probability distributions} $p(X_i)$, rather than $p(X_i | {\bf pa}_i)$. Thus, $p(X_i)$ corresponds to the probability of an attacker to reach a given security condition and, hence, is an indicator of the risk. Using Bayes rule, it is possible to calculate $p(X_i)$ from the product of conditional probability distributions:
\begin{equation}
p(X_i) = \sum_{{\bf X} - X_i} p({\bf X})  = \sum_{{\bf X} - X_i} \prod_{j=1}^n p(X_j | {\bf pa}_j )
\label{eqMarginal}
\end{equation} where ${\bf X} - X_i$ indicates that we sum over all the set of random variables ${\bf X}$ except $X_i$. 

In contrast, for the dynamic analysis of AGs, given evidence of attacks on a set of nodes ${\bf X}_e$, e.g. through SIEM alerts, we need to compute the posterior probability $p(X_i|{\bf X}_e)$, i.e. the probability of an attacker to compromise the node $X_i$ given that we have observed evidence of attack at nodes ${\bf X}_e$. Again, using Bayes rule, we can compute this posterior distribution from the joint probability distribution:
\begin{equation}
p(X_i|{\bf X}_e) = \frac{p(X_i,{\bf X}_e)}{p({\bf X}_e)} = \frac{\sum_{{\bf X} - \{X_i,{\bf X}_e\} } p({\bf X})}{\sum_{{\bf X} - {\bf X}_e} p({\bf X})}
\label{eqPosterior}
\end{equation}

However, the exact calculation of (\ref{eqMarginal}) and (\ref{eqPosterior}) is an NP-Hard problem \cite{koller}. Thus, applying brute force and computing the joint probability distributions to make inference in probabilistic graphical models is not a reasonable approach in terms of computational time and memory. Thus, efficient algorithms, such as Variable Elimination (VE) \cite{dechter} or Junction Tree (JT) \cite{shenoy,shafer}, are necessary even for small graphs. However, the applicability of these techniques is limited, especially when the graphs are dense, demanding a lot of memory to compute the unconditional probabilities. In these cases, approximate inference is a reasonable alternative to enable a tractable estimation of the unconditional and posterior probabilities. Although approximate inference in BNs is also NP-Hard \cite{koller}, efficient techniques like LBP \cite{pearl} allow us to efficiently estimate the unconditional and the posterior probabilities in (\ref{eqMarginal}) and (\ref{eqPosterior}) for large networks. In Section \ref{sec:Inference} we describe how to use LBP for the analysis of BAG.

\subsection{Applying Bayesian Attack Graphs for Security Risk Assessment and Mitigation}
The BAG model can be applied in practice by system administrators to perform  security risk assessment and risk mitigation strategies. For both applications we also distinguish two kind of analysis: \emph{static}, i.e. considering the security posture of the network at rest, and \emph{dynamic}, i.e. when the network is operative and attacks may occur.

For \emph{risk assessment}, the BAG model can be built from the network topology, network reachability, and the results of a vulnerability analysis. Then, we can compute the conditional probability tables for the nodes in the BAG by determining the values of the successful exploitation of vulnerabilities. As shown before, this can be done using the exploitability submetric of the CVSS score. On the resulting BAG we can use exact or approximate inference techniques, such as the JT algorithm or LBP, to perform static risk assessment by computing the unconditional probabilities of all the nodes. These probabilities can be used as risk estimates to detect weak areas in the network and serve as an input for network hardening or static risk mitigation techniques.

For dynamic risk assessment, we recompute the probabilities of the BAG model at run-time taking into account indications that some of the networks components can have been compromised, for example from IDS or SIEM. Thus, the state of the random variables represented by the corresponding nodes for which we observe evidence of compromise are set to $1$ (the true state). Then, the posterior probabilities of the rest of the nodes given the evidence of the compromised nodes are computed. This allows system administrators to dynamically profile the possible attack paths.

On the other hand, the unconditional and posterior probabilities provided by the BAG model can be used to mitigate the potential or existing risks in the network by proposing security countermeasures to effectively reduce the risk of the target nodes to be compromised. In this sense, the network hardening techniques proposed in \cite{noel,albanese,dewri} aim to eliminate or reduce the risk of compromising a target node by proposing a set of countermeasures. In \cite{noel,albanese}, this is achieved using monotonic logic heuristics, whereas in \cite{dewri} the authors also consider a cost model of implementing the countermeasures. However, these techniques are restricted to static risk mitigation. In contrast, \cite{poolsappasit} model risk mitigation as a discrete reasoning problem solved using a genetic algorithm that can be applied for dynamic risk mitigation in combination with the output provided by a BAG model.

\subsection{Example}
\begin{figure}
	\centering
	\includegraphics[width=9cm,height=6cm]{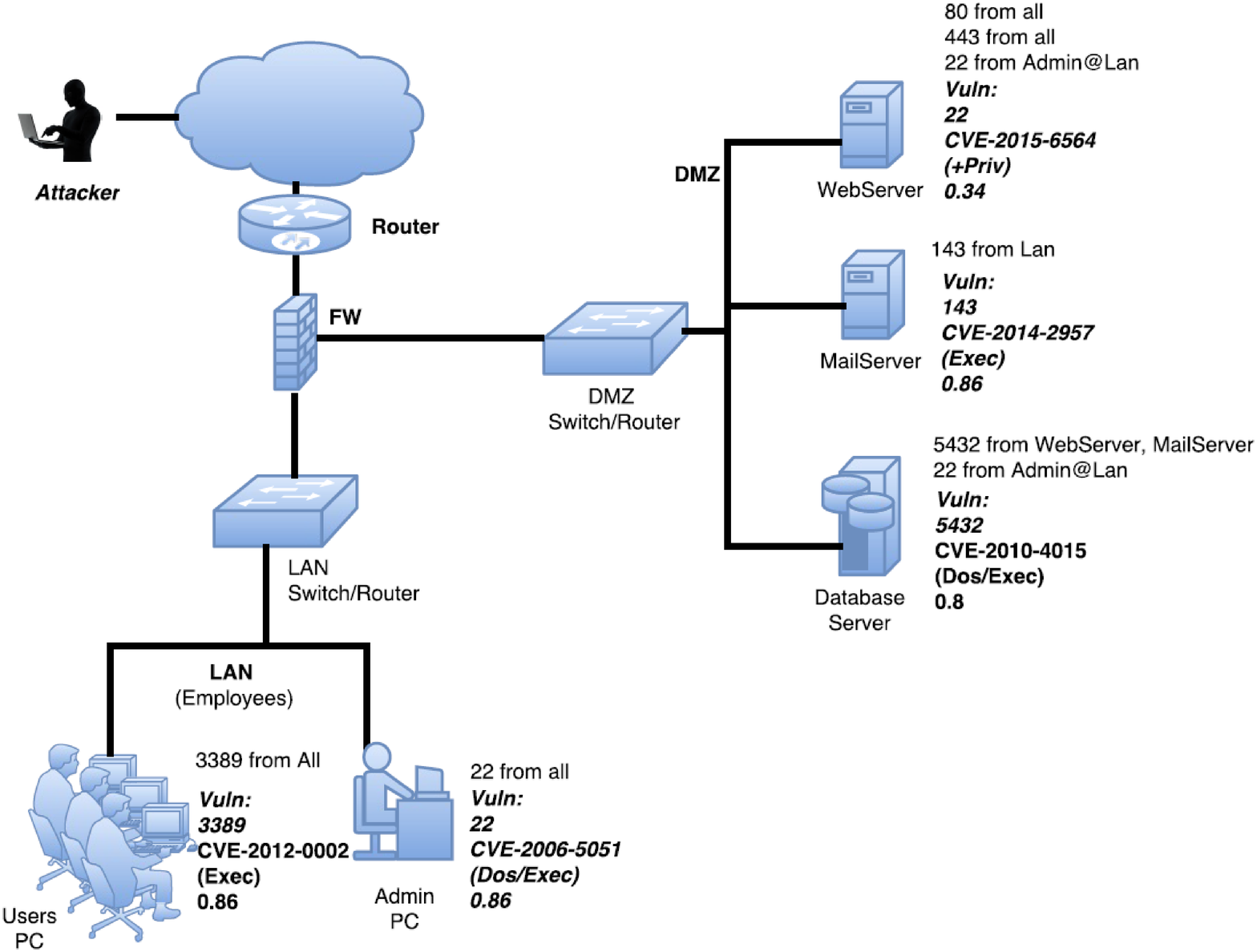} 
	\caption{Example of a Small Network with one LAN, a Mail server, a Web server, and a Database server.}
	\label{UseCase}	
\end{figure}

In this subsection we show an example of analysis of a BAG in the scenario depicted in Fig. \ref{UseCase}, which represents a typical small corporate network. We will also use this example to illustrate and explain the inference algorithms described in Sections \ref{sec:Inference} and \ref{sec:JT}.

In detail, for the network in Fig.~\ref{UseCase} we have an internal LAN for corporate employees, and a DMZ hosting the company's servers, namely, a public Web server, a Mail server, and a local Database server (used to store public and private data). For each node, we have indicated the set of reachable ports, and from which other nodes they are reachable: this set includes those ports open/filtered by the firewall, as well as those open/closed by local firewalls, switches, routers, etc. In addition, we have highlighted some vulnerabilities that might be present on the network nodes. As an example, the Web server can be accessed on port 80 and port 43 by any other node (and also from the Internet), whereas it can be reached on port 22 (SSH server) only from the IP addresses belonging to the ``Admin PC'' node in the LAN\footnote{If not explicitly indicated, it means that any other port is closed.}. Further, we suppose this node has a vulnerability affecting the SSH server (CVE-2015-6564). For each vulnerability, we show in Fig.~\ref{UseCase} over which port it can be exploited (in case the vulnerability is a remote one), the CVE identifier, the type of vulnerability (DoS, elevation of privilege, etc.), and the likelihood of exploiting such a vulnerability. We have based this likelihood on the CVSS Exploitability Subscore, which we have divided by 10. In our particular example, when the corresponding score is 1.0 (which means, an exploit already exists and it is ready/easy to use), we have decided to lower the value to 0.95, since a probability of successful exploitation of 1.0 means that the attacker has already reached the next security state, without necessarily exploiting the vulnerability, which is not true. Finally, we suppose a generic attacker exists that is willing to launch attacks from the Internet. 

\begin{figure}
	\centering
	\includegraphics[width=6.5cm,height=3.7cm]{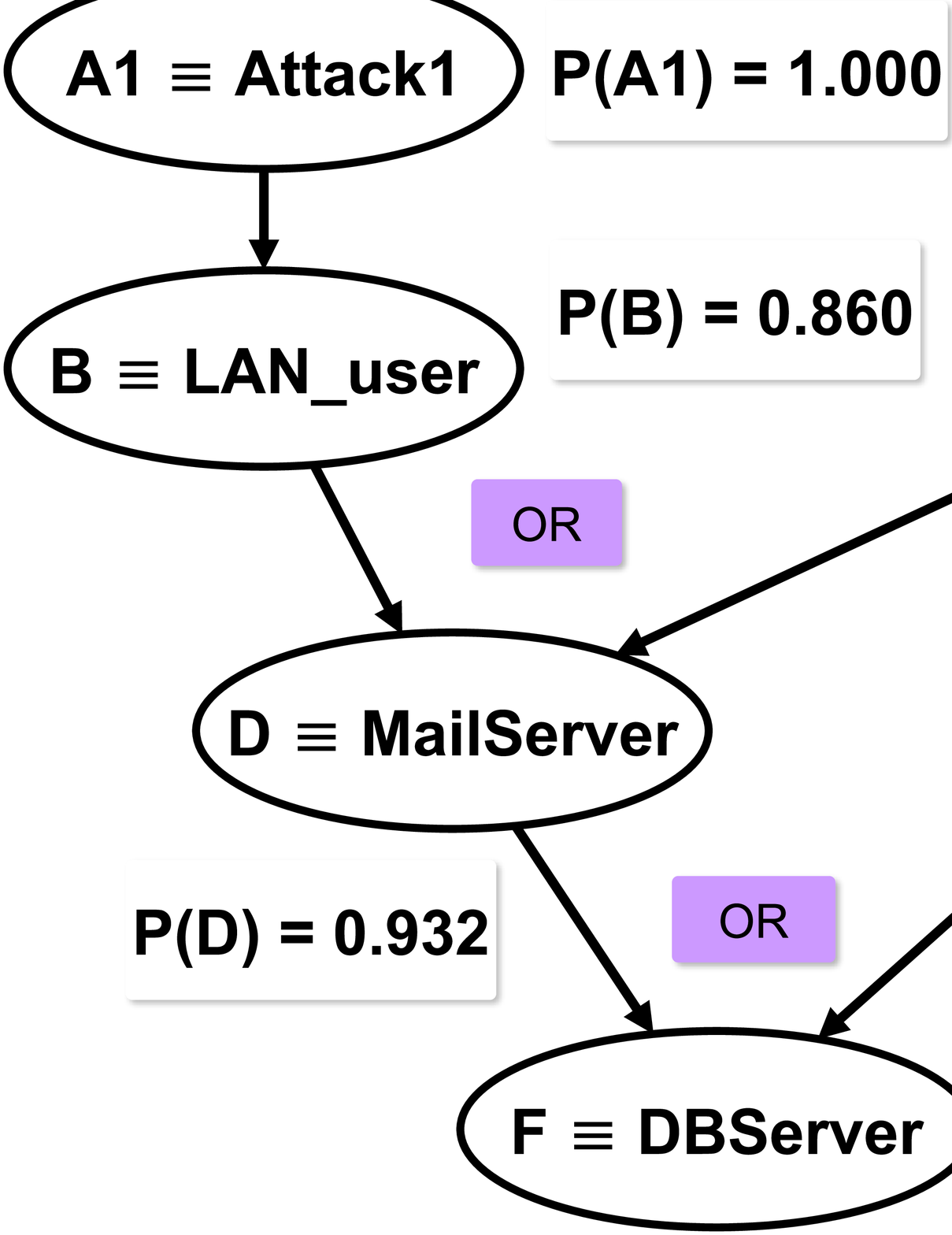} \ \ 
	\includegraphics[width=6.5cm,height=3.7cm]{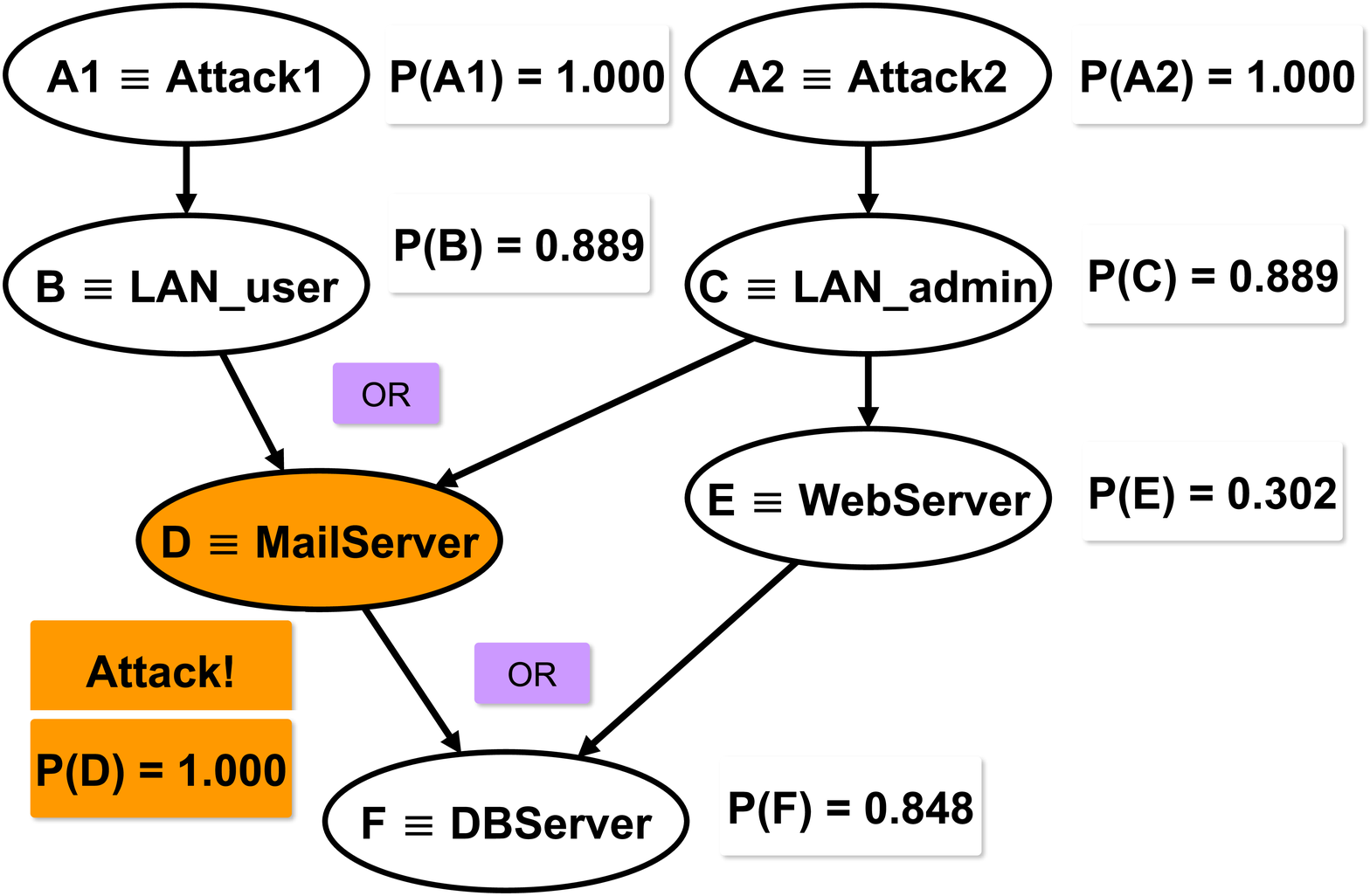} \vspace{0.2cm} \\
	(a) \hspace{7cm}(b)  \\
	\caption{BAG example: (a) Unconditional probabilities for the static analysis. (b) Posterior probabilities when there is evidence of attack at the Mail Server.}
	\label{BAG1}	
\end{figure}

The BAG representation for this example is shown in Fig. \ref{BAG1}. Following the guidelines described before, we have two nodes $A_1$ and $A_2$ that represent the initial attacker's state for the two possible initial attack paths, and the final objective of the attacker is to compromise the Database server (node $F$). With these settings, the joint probability for all the nodes in the BAG can be written as:
\begin{equation}
p(A_1,A_2,B,C,D,E,F) =  p(A_1) \ p(A_2) \ p(B|A_1) \ p(C|A_2) \ p(D|B,C) \ p(E|C) \ p(F|D,E) 
\label{eqJoint}
\end{equation}

In Fig.~\ref{BAG1}.(a) we show the result of the static analysis of the BAG, where we are interested in computing the unconditional probabilities when no evidence of attack is observed. In this case, we observe that the Database server can be compromised with a probability of $79.9\%$ and that this high risk is due to the high probability of compromising the Mail server. We can use this analysis to prioritize patching of the most critical vulnerabilities present in the network, as proposed in \cite{albanese}.

\begin{figure}
	\centering
	\includegraphics[width=6.5cm,height=3.7cm]{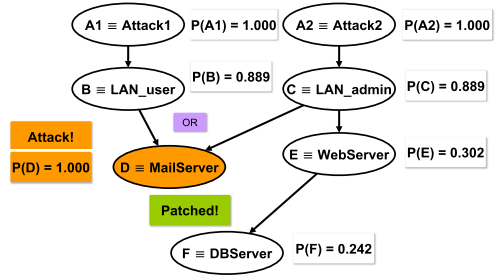}
	\caption{Updated posterior probabilities when the vulnerability from the Mail Server to the Data Base server has been patched in the BAG example in Fig.~\ref{BAG1}. }
	\label{BAG2}	
\end{figure}

An example of dynamic analysis of the BAG is shown in Fig.~\ref{BAG1}.(b), where we consider that the alert correlation system triggers an alarm on the Mail server. For the sake of clarity, we have considered here a perfect behaviour of the alert correlation system. Fig.~\ref{BAG1}.(b) shows the posterior probabilities for all the network nodes given the evidence of compromise at node $D$. As expected, we observe that the risk of compromising the Database server increases to $84.8\%$. When reasoning about the potential attack path allowing the attacker to compromise the Mail server, the posterior probability of both nodes $B$ and $C$ is the same - in this example the paths are equally likely. 

In Fig.~\ref{BAG2} we show the updated posterior probabilities when the system administrator has patched the vulnerability from the Mail Server to the Data Base Server, so that the corresponding attack path has been removed in the BAG representation. To recompute the posterior probabilities, we just need to update the conditional probability table for node $F$ (the Data Base server) by considering that the probability of successful exploitation of the vulnerability from the Mail Server is zero. Then, we recompute the posterior probabilities given the evidence of attack in the Mail Server. As shown in Fig.~\ref{BAG2}, this countermeasure reduces the risk of the Data Base server being compromised from $84.8\%$ to $24.2\%$. In the next section we describe how these probabilities can be efficiently computed.

\section{Scalable Inference on Bayesian Attack Graphs} \label{sec:Inference}
In this section we introduce two efficient techniques to make inference in BAGs: Belief Propagation (BP) and Loopy Belief Propagation (LBP). Both algorithms are based on message passing schemes that allow to compute the unconditional probabilities of the nodes in BNs or Markov Random Fields (MRFs). While BP is a method for exact inference, it is restricted to tree or polytree graph structures. On the contrary, LBP is an approximate inference technique that can be applied to graphs with loops. 


%

\subsection{Belief Propagation} \label{sec:BP}
As mentioned earlier, BP allows to compute the unconditional probabilities in a BN or a MRF when the graph is a tree or a polytree  \cite{pearl82,pearl}. Although this structure is restrictive for AGs in general, BP can be applied for inference on Attack Trees \cite{schneier}. Furthermore, for graphs with a general structure we can use the JT algorithm as an extension of BP, as we describe in Section \ref{sec:JT}. To describe the BP algorithm, we introduce the concept of \emph{factor graphs} using a formulation similar to that given in \cite{bishop}. BNs and MRFs allow us to express the joint probability distribution of a set of random variables as a product of factors over subsets of those variables. Factor graphs make this factor decomposition explicit by introducing additional nodes for the factors themselves, in addition to those representing the variables, thus resulting in bipartite graphs.

\begin{figure}
	\centering
	\includegraphics[width=4.5cm,height=3.5cm]{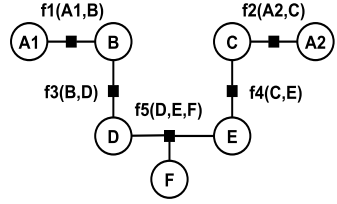}
	\caption{Factor Graph representation for the BAG in Fig.~\ref{BAG1} removing the attack path from the LAN admin to the Mail server.}
	\label{fgBP}	
\end{figure}

Referring to the example shown in Fig.~\ref{BAG1}, if we remove the attack path from the LAN admin to the Mail server (for example, by patching the vulnerability), the resulting AG is a tree and we can use BP for inference. For this reduced BAG, the joint probability distribution can be factorised as:
\begin{equation}
p(A_1,A_2,B,C,D,E,F) = \prod_{i=1}^5 f_i ({\bf X_i})
\label{eqBP1}
\end{equation} where the corresponding factors $f_i ({\bf X_i})$ are:
\begin{equation}
\begin{split}
f_1(A_1,B) & = p(A_1) \ p(B|A_1) \\
f_2(A_2,C) & = p(A_2) \ p(C|A_2) \\
f_3(B,D) & = p(D|B) \\
f_4(C,E) & = p(E|C) \\
f_5(D,E,F) & = p(F|D,E) \\
\end{split}
\label{eqBP2}
\end{equation}
Note that several factor graph representations may exist for a given BN (or MRF) \cite{bishop}. However, the selection of a specific representation does not significantly impact the performance of BP. In our example, the corresponding factor graph according to (\ref{eqBP2}) is shown in Fig.~\ref{fgBP}.

BP works by passing real valued functions called messages among the neighbouring nodes in the graph. Since factor graphs are bipartite, there are two possible types of messages: From variable to factor, and from factor to variable. The message from a variable $X_i$ to a factor $f_j$ in the neighbourhood of $X_i$ is given by:
\begin{equation}
\mu_{X_i, f_j} (X_i) = \prod_{f_k \in \{{\bf F}_i - f_j\}} \mu_{f_k, X_i} (X_i)
\label{eqBP3}
\end{equation} where $\mu_{f_k, X_i} (X_i)$ are the messages from the factor nodes in the neighbourhood of $X_i$ except $f_j$. On the other hand, the message from a factor node $f_i$ to a variable node $X_j$ in the neighbourhood of $f_i$ is calculated as:
\begin{equation}
\mu_{f_i, X_j} (X_j) = \sum_{X_k \in {\bf X}_s} f_i(X_j,{\bf X}_s) \prod_{X_k \in {\bf X}_s} \mu_{X_k, f_j} (X_k)
\label{eqBP4}
\end{equation} where ${\bf X}_s$ is the set of variable nodes in the neighbourhood of $f_i$ except $X_j$.

When a variable $X_i$ is a leaf node, the corresponding messages to the factors in its neighbourhood are equal to one, i.e. $\mu_{X_i, f_j} (X_i) = 1$. On the contrary, if a factor $f_i$ is a leaf node, the message to a variable node in its neighbourhood is given by $\mu_{f_i, X_j} (X_j) = \sum_{X_k \in {\bf X}_s} f_i(X_j,{\bf X}_s)$. 

To compute the unconditional probabilities for all the nodes in the graph, BP needs to compute all the messages from all node variables to their corresponding factors and vice versa. BP proceeds starting from the leaf nodes (either variable or factor nodes) and propagates the messages across the graph such that a variable node $X_i$ cannot send a message to a factor $f_j$ until $X_i$ receives all messages from its neighbouring factors except $f_j$. The same applies when sending messages from factors to variable nodes. For example, in the factor graph in (\ref{fgBP}), we cannot send a message from factor $f_5$ to variable $F$ until $f_5$ receives a message from variables $D$ and $E$.

Once all messages are computed, the unconditional probability for a node $X_i$, when the graph is a BN\footnote{For MRFs we need to include a normalization factor.}, can be calculated as:
\begin{equation}
p(X_i) = \prod_{f_j \in {\bf F}_i} \mu_{f_j,X_i} (X_i)
\label{eqBP5}
\end{equation} where ${\bf F}_i$ are the factor nodes in the neighbourhood of $X_i$.  Therefore, BP can efficiently calculate all the marginal probabilities by computing all the messages once and storing them. 

For the dynamic analysis, when we observe new evidence of compromise in some nodes, we only need to recompute the factors that depend on the variables that have changed in order to obtain the posterior probability on all the nodes of the network given the evidence of compromise. Further details are explained in \cite{koller}. Finally, the details about the computational complexity of BP will be discussed in Section \ref{sec:JT} along with the corresponding discussion for the complexity of JT, as BP can be considered as a particular case of JT.

\subsection{Loopy Belief Propagation} \label{sec:LBP}
LBP is a simple extension of BP \cite{pearl} applied to graphs (BNs or MRFs) that contain loops. The difference is that, in the presence of loops, the results of LBP are approximate estimates of the unconditional probabilities of the nodes in the graph.

\begin{figure}
	\centering
	\includegraphics[width=4.5cm,height=3.2cm]{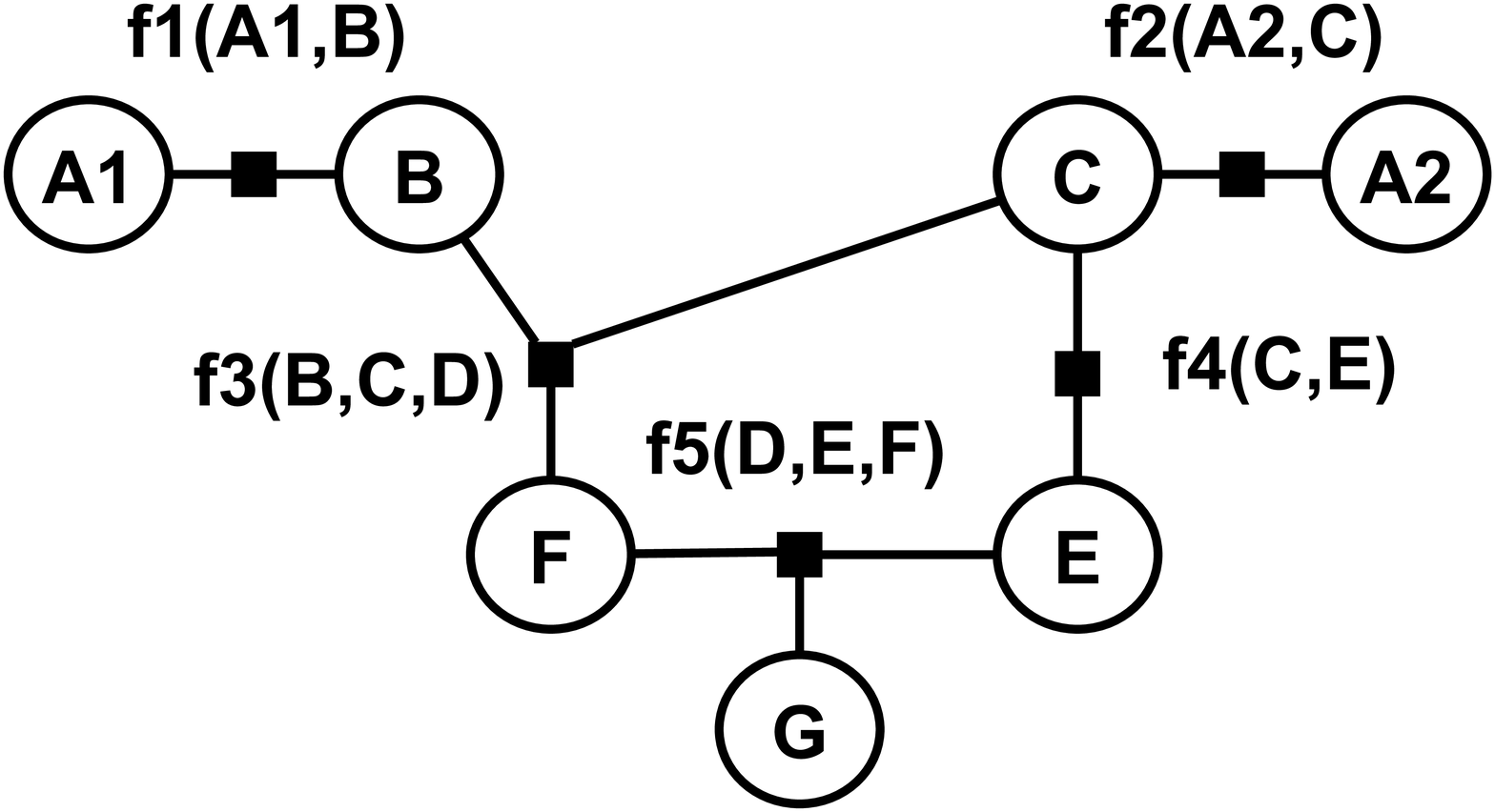}
	\caption{Factor Graph representation for the BAG in Fig.~\ref{BAG1}.}
	\label{fgLBP}	
\end{figure}

LBP uses the same factor graph representation as BP. In Fig.~\ref{fgLBP} we show the corresponding factor graph representation for the BAG depicted in Fig.~\ref{BAG1}, where we observe that there is one loop due to the attack path from the LAN admin node to the Mail server. The corresponding factors $f_i ({\bf X_i})$ are the same as in (\ref{eqBP2}) except for $f_3$, which in this case also depends on variable $C$, so that $f_3(B,C,D) = p(D|B,C)$.

There are two possible implementations of LBP according to how messages are computed, namely \emph{Sequential LBP} (S-LBP) and \emph{Parallel LBP} (P-LBP) \cite{murphy} (they sometimes are also referred as asynchronous and synchronous LBP respectively). For S-LBP we iteratively compute the messages in (\ref{eqBP3}) and (\ref{eqBP4}) following some arbitrary schedule, until the unconditional probability estimates obtained with (\ref{eqBP5}) converge or until a maximum number of iterations has been reached. Although there is no restriction on the order in which we update the messages (\ref{eqBP3}) and (\ref{eqBP4}), and the beliefs (\ref{eqBP5}), depending on the structure of the graph, there are some scheduling techniques that can be applied to favour convergence and reduce the time to achieve it \cite{koller}. In Algorithm \ref{alg:SLBP} we have detailed the steps to compute and update the messages in S-LBP.

\begin{algorithm}[t]
\SetAlgoNoLine
\KwIn{Set of nodes ${\bf X}$, set of factors ${\bf F}$.}
\KwOut{Messages from nodes to factors, $\mu_{X_i, f_j}$, and from factors to nodes, $\mu_{f_i, X_j}$.}
\For{each node $X_i$ in ${\bf X}$
    }{
      \For{each factor $f_j$ in ${\bf F}_i$ (the neighbourhood of $X_i$)
      }{
         $\mu_{X_i, f_j} (X_i) \leftarrow 1$\;
      }       
      }
 \For{each factor $f_i$ in ${\bf F}$
    }{
      \For{each node $X_j$ in ${\bf X}_s$ (the neighbourhood of $f_i$)
      }{
         $\mu_{f_i, X_j} (X_j) \leftarrow \sum_{X_k \in {\bf X}_s} f_i(X_j, {\bf X}_s)$\;
      }      
      
      }      
\caption{Initialize messages for LBP}
\label{alg:Initialize}
\end{algorithm}

\begin{algorithm}[t]
\SetAlgoNoLine
\KwIn{Set of nodes ${\bf X}$, set of factors ${\bf F}$, convergence tolerance $\epsilon$, maximum number of iterations $max\_iter$.}
\KwOut{Unconditional probabilities $p(X_i)$ for all the nodes in ${\bf X}$.}
${\text iter} = 0$\;
Initialize messages according to Algorithm \ref{alg:Initialize}\;   
      
\Repeat{$\left( \sum_{i=1}^{N} | p(X_i) - p(X_i)^{\text{old}} | < \epsilon \right)$ OR $\left( {\text iter} \geq {\text max\_iter} \right)$}{
        ${\text iter} \leftarrow {\text iter} + 1$\;
        \For{each node $X_i$ in ${\bf X}$
    }{
      $p(X_i)^{{\text old}} \leftarrow p(X_i)$\;     
      }
      
      \For{each node $X_i$ in ${\bf X}$
    }{
      \For{each factor $f_j$ in ${\bf F}_i$ (the neighbourhood of $X_i$)
      }{
         $\mu_{X_i, f_j} (X_i) = \prod_{f_k \in \{{\bf F}_i - f_j\}} \mu_{f_k, X_i} (X_i)$\;
      }      
      
      }
      \For{each factor $f_i$ in ${\bf F}$
    }{
      \For{each node $X_j$ in ${\bf X}_s$ (the neighbourhood of $f_i$)
      }{
         $\mu_{f_i, X_j} (X_j) = \sum_{X_k \in {\bf X}_s} f_i(X_j,{\bf X}_s) \prod_{X_k \in {\bf X}_s} \mu_{X_k, f_j} (X_k)$\;
      }      
      
      }
     \For{each node $X_i$ in ${\bf X}$
    }{      
	  $p(X_i) = \prod_{f_j \in {\bf F}_i} \mu_{f_j,X_i} (X_i)$\;      
      }
      }
\caption{Sequential LBP}
\label{alg:SLBP}
\end{algorithm}

In contrast, P-LBP updates all the messages for all factors and variable nodes at the same time, using the values of the messages at the previous iteration. Thus, at iteration $t$, we first update the messages from nodes to factors. The update equation for the message from a node $X_i$ to a factor $f_j$ can be written as:
\begin{equation}
\mu_{X_i, f_j}^{(t)} (X_i) = \prod_{f_k \in \{{\bf F}_i - f_j\}} \mu_{f_k, X_i}^{(t - 1)} (X_i)
\label{eqLBP1}
\end{equation} In a second step we update the messages from factors to variable nodes where the equation the message from factor $f_i$ to node $X_j$ is given by:
\begin{equation}
\mu_{f_i, X_j}^{(t)} (X_j) = \sum_{X_k \in {\bf X}_s} f_i(X_j,{\bf X}_s) \prod_{X_k \in {\bf X}_s} \mu_{X_k, f_j}^{(t)} (X_k)
\label{eqLBP2}
\end{equation} Finally, we compute the new estimates of the marginal probabilities with (\ref{eqBP5}) with the updated messages obtained using (\ref{eqLBP2}). As in the previous case, the algorithm is repeated until the unconditional probabilities converge or a maximum number of iterations is reached. The details of P-LBP are shown in Algorithm \ref{alg:PLBP}. 

For the first iteration in both S-LBP and P-LBP we initialize the messages from nodes to factors to $1$. The messages from a factor $f_i$ to a node $X_j$ are initialized as $\mu_{f_i, X_j}(X_j) = \sum_{X_k \in {\bf X}_s} f_i(X_j, {\bf X_s})$. This procedure is described in Algorithm \ref{alg:Initialize}. 

\begin{algorithm}[t]
\SetAlgoNoLine
\KwIn{Set of nodes ${\bf X}$, set of factors ${\bf F}$, convergence tolerance $\epsilon$, maximum number of iterations $max\_iter$.}
\KwOut{Unconditional probabilities $p(X_i)$ for all the nodes in ${\bf X}$.}
${\text iter} = 0$\;
Initialize messages according to Algorithm \ref{alg:Initialize}\;       
      
\Repeat{$\left( \sum_{i=1}^{N} | p(X_i) - p(X_i)^{\text{old}} | < \epsilon \right)$ OR $\left( {\text iter} \geq {\text max\_iter} \right)$}{
        ${\text iter} \leftarrow {\text iter} + 1$\;
        \For{each node $X_i$ in ${\bf X}$
    }{
      $p(X_i)^{{\text old}} \leftarrow p(X_i)$\;
      \For{each factor $f_j$ in ${\bf F}_i$ (the neighbourhood of $X_i$)
      }{
         $\mu_{X_i, f_j}^{{\text old}} (X_i) \leftarrow \mu_{X_i, f_j} (X_i)$\;
      }      
      
      }
      \For{each factor $f_i$ in ${\bf F}$
    }{
      \For{each node $X_j$ in ${\bf X}_s$ (the neighbourhood of $f_i$)
      }{
         $\mu_{f_i, X_j}^{{\text old}} (X_j) \leftarrow \mu_{f_i, X_j} (X_j)$\;
      }      
      
      }
      \For{each node $X_i$ in ${\bf X}$
    }{
      \For{each factor $f_j$ in ${\bf F}_i$
      }{
         $\mu_{X_i, f_j} (X_i) = \prod_{f_k \in \{{\bf F}_i - f_j\}} \mu_{f_k, X_i}^{{\text old}} (X_i)$\;
      }      
      
      }
      \For{each factor $f_i$ in ${\bf F}$
    }{
      \For{each node $X_j$ in ${\bf X}_s$
      }{
         $\mu_{f_i, X_j} (X_j) = \sum_{X_k \in {\bf X}_s} f_i(X_j,{\bf X}_s) \prod_{X_k \in {\bf X}_s} \mu_{X_k, f_j}^{{\text old}} (X_k)$\;
      }      
      
      }
     \For{each node $X_i$ in ${\bf X}$
    }{      
	  $p(X_i) = \prod_{f_j \in {\bf F}_i} \mu_{f_j,X_i} (X_i)$\;      
      }
      }
\caption{Parallel LBP}
\label{alg:PLBP}
\end{algorithm}

According to \cite{koller}, S-LBP usually works better than P-LBP, although they require scheduling messages in a guided way. However, the experimental results on synthetic BAGs, presented in Section \ref{sec:Experiments}, show a similar behaviour for both implementations in terms of accuracy. 

\begin{figure}
	\centering
	\includegraphics[width=6.5cm,height=3.7cm]{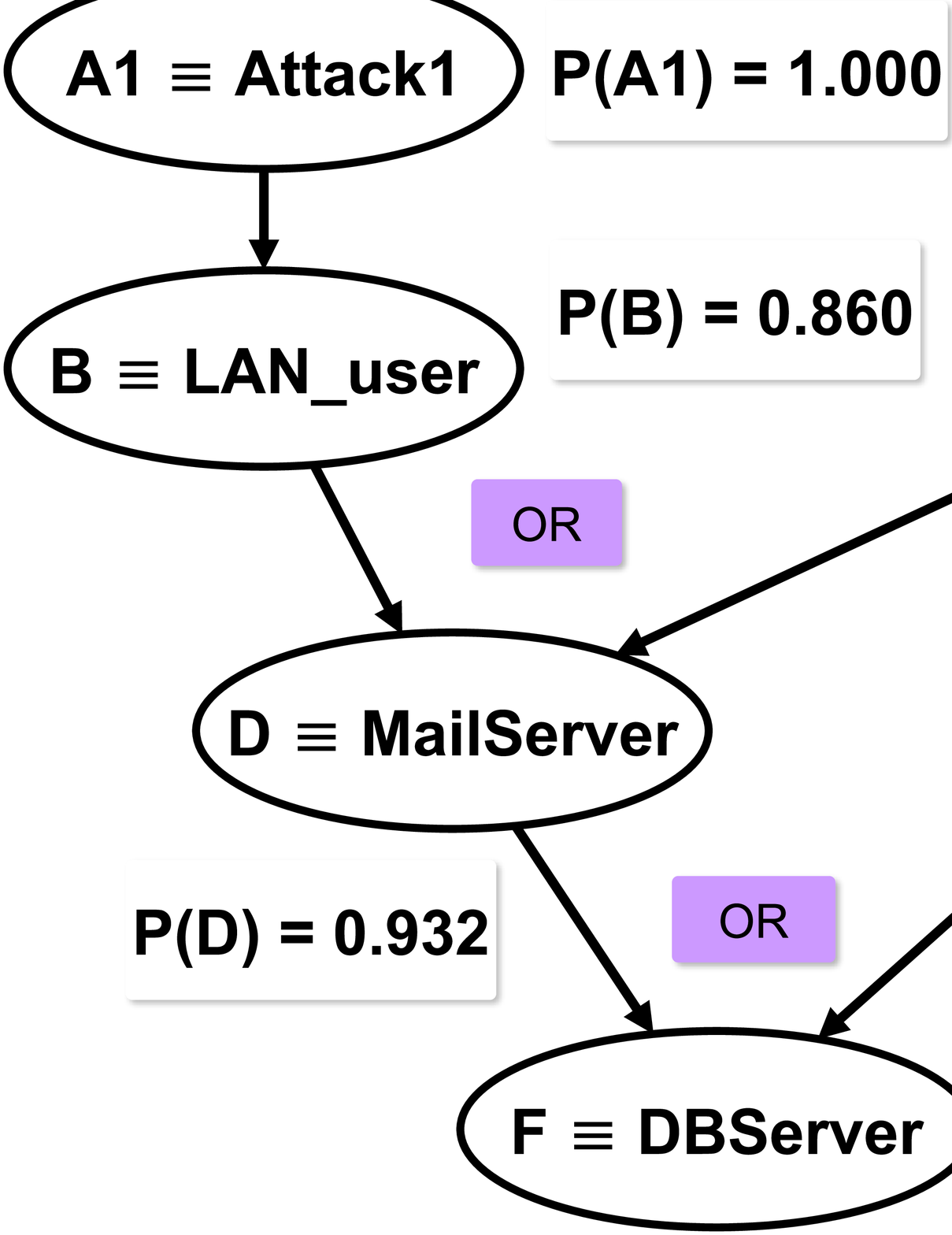}
	\caption{Estimation of the unconditional probabilities provided by LBP for the BAG in Fig.~\ref{BAG1}.(a)}
	\label{BAG_LBP}	
\end{figure}

In Fig.~\ref{BAG_LBP} we show the corresponding estimates for the unconditional probabilities in the BAG depicted in Fig.~\ref{BAG1}.(a) when there is no evidence of attack. It can be observed that there is no difference (at least in the $3$ first decimals) between the true and the estimated probabilities for nodes $A$-$E$ and that, for node $F$, the marginal probability estimated with LBP\footnote{In this case both S-LBP and P-LBP provide the same result.} is $0.805$ while the true probability is $0.799$. When evidence of attack is observed at the Mail server (node $D$), shown in Fig.~\ref{BAG1}.(b),  the LBP unconditional probability estimates match the exact probabilities. The reason for this exact result is that, given the evidence of attack, $D$ can be considered as a deterministic node. This splits the graph into two tree structures with the remaining unobserved nodes: $\{A_1,B\}$ and $\{A_2,C,E,F\}$ and the message passing scheme produces exact results for the two trees.

One of the drawbacks of LBP is that, in general, convergence is not guaranteed. \cite{weiss} show that LBP converges for graphs with a single loop and derives an analytical relationship between LBP probability estimates and the true unconditional probabilities. For more general graphs, \cite{mooij,ihler2005} present sufficient conditions on the convergence of LBP based on the concept of $\alpha$-contractions. However, applying the corresponding analysis is, in general, a difficult task \cite{koller}. It is also important to note that convergence does not mean correctness, i.e. LBP convergence does not imply that the unconditional probability estimates are accurate. However, the empirical study proposed in \cite{murphyLBP} shows that usually, when LBP converges, the approximate marginal probabilities are close to the exact values.

One simple way to favour convergence is to use \emph{damping}. Hence, the update of the messages from a variable node $X_i$ to a factor $f_j$ have the form:
\begin{equation}
\hat{\mu}_{X_i, f_j}^{(t)} (X_i) = \alpha \mu_{X_i, f_j}^{(t)} (X_i) + (1 - \alpha) \mu_{f_k, X_i}^{(t - 1)} (X_i)
\label{eqLBP3}
\end{equation} so that the damped message $\hat{\mu}_{X_i, f_j}^{(t)} (X_i)$ is a convex combination of the message update at iteration $t$ and the message at iteration $t-1$, where the damping factor $\alpha$ is a positive value smaller than $1$. The damped update for the messages from factors to variable nodes is analogous. This technique can be applied for both S-LBP and P-LBP. In the experiments, in Section \ref{sec:Experiments}, we analyse the effect of damping for the convergence and accuracy of LBP. 

It is also possible to make LBP converge to a local minimum using double loop algorithms \cite{yuille,welling}. Unfortunately, these techniques are slow and complicated and their accuracy is often worse than the standard LBP \cite{murphy}, since they are prone to converge to poor local minima. Similarly, other techniques, such as the mean field approximation, have been proved to converge but are usually less accurate than LBP, since the non-convexity of the mean field objective function leads to poor solutions \cite{weiss2001}.

\section{Junction Tree} \label{sec:JT}
The existing literature on BAGs has focused on the use of exact inference techniques for the static and dynamic analysis of AGs. \cite{liu} propose to use Variable Elimination as a mechanism to compute the unconditional probabilities in the graph. \cite{poolsappasit} propose to use forward-backward propagation, however this technique can only be applied when the graph is a chain \cite{rabiner,murphy}, which is not the case for most AGs. Finally, \cite{luis} propose to use the JT algorithm to efficiently compute the unconditional probabilities enabling the static and dynamic analysis of AGs with hundreds of nodes. The experimental evaluation in \cite{luis} shows the advantages of the JT compared to the Variable Elimination algorithm proposed in \cite{liu} in terms of computational complexity and memory required. Thus, in this section we describe the JT algorithm, as the state-of-the-art technique to perform exact inference for the static and dynamic analysis of BAGs and serves as benchmark for the comparison with LBP.

The JT or clique tree algorithm is an extension of BP, for exact inference, that can be applied on BNs or MRFs with a general structure. In this case, BP's message scheme is applied to a tree structure where the nodes represent clusters of the random variables in the graph. There are two main variants for the JT: The Shenoy-Shafer algorithm \cite{shenoy,shafer} and the Hugin algorithm \cite{lauritzen}. Although both techniques rely on the same principles, they differ in the way the messages are computed. In the following, we will describe the Shenoy-Shafer method which uses the same message passing scheme as BP.

The first step of JT is to create a cluster graph with a tree structure from the initial BN (or MRF). This cluster graph (or clique tree) can be considered an extension of factor graphs with clusters of several random variables between two factors. In this case, one random variable can appear in more than one cluster node. However, the cluster graph needs to satisfy the \emph{running intersection property}: if a random variable $X_i$ appears in two cluster nodes, $X_i \in C_j$ and $X_i \in C_k$, then $X_i$ also appears in each cluster node in the unique path existing between $C_j$ and $C_k$ in the clique tree. 

In the case of BNs, to create a clique tree we first need to moralize the graph, i.e. make the graph undirected and add a link between the nodes that have a common child (this step is not needed for MRFs). The moral graph is then triangulated to obtained a chordal graph, i.e. one in which every minimal loop in the graph is of length three. Finally, each maximal clique in the chordal graph is a cluster node in the clique tree. As shown in \cite{koller}, we can create the cluster graph with the VE algorithm \cite{dechter}. 

\begin{figure}
	\centering
	\includegraphics[width=6.7cm,height=1.8cm]{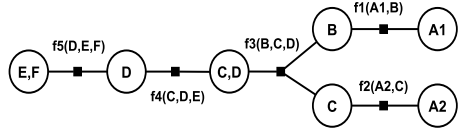}
	\caption{Factor Graph representation for clique tree obtained from the BAG in Fig.~\ref{BAG1}.}
	\label{fgJT}	
\end{figure}

For the BAG example in Fig.~\ref{BAG1},  we show in Fig.~\ref{fgJT} a factor graph representation of the corresponding clique tree obtained using VE. For this factor graph representation, the assignment of the different terms in (\ref{eqJoint}) to the factors in Fig.~\ref{fgJT} is:
\begin{equation}
\begin{split}
f_1(A_1,B) & = p(A_1) \ p(B|A_1) \\
f_2(A_2,C) & = p(A_2) \ p(C|A_2) \\
f_3(B,C,D) & = p(D|B) \\
f_4(C,D,E) & = p(E|C) \\
f_5(D,E,F) & = p(F|D,E) \\
\end{split}
\label{eqJT1}
\end{equation}

With the factor graph representation of the clique tree we can calculate the unconditional probabilities using the same message passing scheme as in BP. The difference is that, in this case, the scopes of the messages given in equations (\ref{eqBP3}) and (\ref{eqBP4}) depend on multiple random variables rather than just one, as in the case of BP.

Once all the messages are computed, the unconditional joint probability for the variables in a cluster node ${\bf X}_s$ (provided that the graph is a BN) is calculated as:
\begin{equation}
p({\bf X}_s) = \prod_{f_j \in {\bf F}_s} \mu_{f_j,{\bf X}_s}({\bf X}_s)
\label{eqJT2}
\end{equation} where ${\bf F}_s$ are the factor nodes in the neighbourhood of the cluster node ${\bf X}_s$. To calculate the marginal probability for one random variable $X_i$ in the cluster ${\bf X}_s$, we just sum over the rest of the variables in ${\bf X}_s$:
\begin{equation}
p(X_i) = \sum_{X_j \in \{ {\bf X}_s - X_i \}} p({\bf X}_s)
\label{eqJT4}
\end{equation} 

Evidence of compromise can be easily included when using JT, in the same way as in BP. Further details can be found in \cite{koller}. The computational complexity of JT is exponential in the scope of the biggest factor in the clique tree. Concretely, if all the variables in the graph are discrete and have $K$ possible values each (in our case, $K = 2$), JT scales in time and space as ${\cal O} (|F| K^s)$, where $|F|$ is the number of factors and $s$ is the size of the scope of the largest factor in the clique tree (3 in the example in Fig.~\ref{fgJT}). Moreover, the computational complexity of applying VE algorithm to build the clique tree is also exponential. This can limit the application of JT for large graphs, although it depends on the structure of the graph, as we will show in the experiments. The scalability for BP and LBP is similar to that of JT, i.e. they scale in time and space as ${\cal O} (|F| K^s)$. However, since BP and LBP do not cluster variables, the factors are usually smaller, so we expect to have smaller $s$.

\section{Experiments} \label{sec:Experiments}

%
%

In this section we present an experimental evaluation comparing the accuracy and performance of sequential and parallel implementations of LBP with that of the JT algorithm used in \cite{luis}. In detail, we first analyse the accuracy of LBP when computing the unconditional and posterior probabilities for the static and dynamic analysis of AGs, and then compare the time required to estimate these probabilities with that required by JT. This allows us to determine if LBP is a suitable alternative for tractable analysis of large AGs, and if the risk estimates are sufficiently accurate to help administrators propose risk mitigation strategies.

For JT we have used VE to build the clique tree, selecting the elimination ordering according to the \emph{min-weight} heuristic \cite{koller}. As shown in \cite{luis}, the elimination ordering has an impact on the performance of the algorithm, reducing the memory required and the time to compute the unconditional probabilities. For S-LBP and P-LBP we have used a tolerance threshold of $10^{-3}$ for the convergence of the algorithm, i.e. we assume that the algorithm has converged if the biggest change in the unconditional probabilities is less than $10^{-3}$. We have used the Bayes Net toolbox for Matlab\footnote{https://github.com/bayesnet/bnt} for all the algorithms.

To provide a comprehensive evaluation with different graph sizes, different possible network topologies, and interdependencies, we have generated synthetic AGs in the experiments. Note that, currently, there are no collections of AGs of similar variety obtained empirically from real systems; in fact, no collections of empirically obtained AGs exist in the public domain at all. Furthermore, from the examples reported in the literature it is hard to determine the typical graph structures of AGs, e.g. for large corporate networks. We expect these graph structures to vary significantly since they depend on the network topology and the distribution and type of vulnerabilities across the network components. For these reasons, we have used the structures proposed in \cite{luis} to generate the synthetic AGs: \emph{pseudo-random graphs}, where we control the in-degree of the nodes, which is related to the number of vulnerabilities that a node in the network can have; and \emph{cluster graphs}, which model scenarios with different subnetworks that are weakly connected. Then, for each subnetwork, we generate the AG with a pseudo-random structure.

\begin{figure}
	\centering
	\psfrag{1}{{\scriptsize 1}}
	\psfrag{2}{{\scriptsize 2}}
	\psfrag{3}{{\scriptsize 3}}
	\psfrag{4}{{\scriptsize 4}}
	\psfrag{A}{{\scriptsize 5}}
	\psfrag{6}{{\scriptsize 6}}
	\psfrag{7}{{\scriptsize 7}}
	\psfrag{8}{{\scriptsize 8}}
	\psfrag{9}{{\scriptsize 9}}
	\psfrag{B}{{\scriptsize 10}}
	\psfrag{perc}{{\scriptsize \%}}
	\psfrag{CVSS}{{\scriptsize CVSS score}}
	\includegraphics[width=7cm,height=4.3cm]{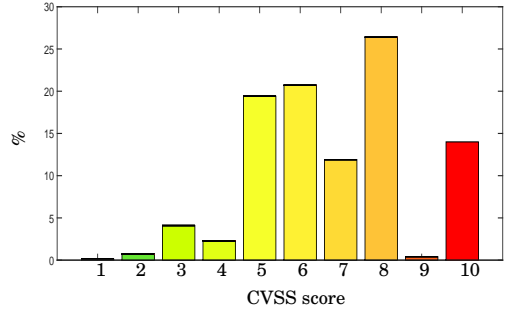}
	\caption{Distribution of the CVSS scores \cite{cvssDistr}.}
	\label{figCVSS}	
\end{figure}

The values for the probabilities of successful exploitation of vulnerabilities are drawn at random from the distribution of CVSS scores extracted from \cite{cvssDistr}. We normalize the scores dividing them by $10$. In Fig.~\ref{figCVSS} we show the distribution of CVSS scores. The value of these probabilities can have an impact in the accuracy of the unconditional probability estimates, and in the convergence of the algorithms. Although we think that the exploitability submetric of CVSS scores is a better indicator of the difficulty of exploiting a vulnerability, it is difficult to get the distribution of this subscore, and so we have used the distribution of the whole CVSS score instead. Finally, since we do not have data to estimate the error probability of the alert systems, and their accuracy changes in time and with the topology of the network, we have considered in our experiments that the alert system does not trigger false alarms. Although we recognise that the false alarm rate is usually very high, we consider that these will be processed and diagnosed before being considered as evidence of compromise. In this sense, we can consider different approaches to reduce and simplify the analysis of IDS alerts: On one side, we can use filtering techniques to reduce the number of alerts to be inspected by discarding those likely to be false positives \cite{CotroneoFGCS,Spathoulas}. On the other hand, we can use correlation systems to identify and cluster alerts pertaining to the same event \cite{Raftopoulos2011,PecchiaSRDS}. We can also reduce the number of false positives with SIEM by combining multiple sources of information (e.g., IDS) to flag potential breaches and by ensuring that the event has been discovered by other monitoring components in the networks \cite{1366134}.

For the experimental evaluation we start analysing the accuracy, convergence, and scalability of S-LBP and P-LBP for the pseudo-random AGs. Then, following a similar treatment we present the experimental results on cluster AGs.

\begin{table}
\tbl{Average RMSE plus/minus one standard deviation for P\-LBP on pseudo-random BAGs with $m=3$ and $40$ nodes varying the damping factor $\alpha$ and the probability of having AND-type conditional probability tables, p(AND). \label{tabAcc}}{
\begin{tabular}{|c|c|c|c|c|c|}
\hline
p(AND) & $\alpha = 0.0$ & $\alpha = 0.1$ & $\alpha = 0.3$ & $\alpha = 0.5$ \\
\hline
\hline
$0.0$ & $0.0315 \pm 0.0223$ & $0.0272 \pm 0.0187$ & $0.0291 \pm 0.0194$ & $0.0266 \pm 0.0175$ \\
\hline
$0.2$ & $0.0218 \pm 0.0141$ & $0.0222 \pm 0.0143$ & $0.0266 \pm 0.0206$ & $0.0276 \pm 0.0183$ \\
\hline
$0.5$ & $0.0182 \pm 0.0103$ & $0.0150 \pm 0.0067$ & $0.0165 \pm 0.0123$ & $0.0175 \pm 0.0083$ \\
\hline
$0.8$ & $0.0081 \pm 0.0069$ & $0.0108 \pm 0.0097$ & $0.0131 \pm 0.0100$ & $0.0112 \pm 0.0064$ \\
\hline
$1.0$ & $0.0081 \pm 0.0058$ & $0.0069 \pm 0.0047$ & $0.0107 \pm 0.0096$ & $0.0089 \pm 0.0085$ \\
\hline
\end{tabular}}
\end{table}

\subsection{BAGs with pseudo-random structure}
\begin{figure}
	\centering
	\includegraphics[width=12cm,height=8cm]{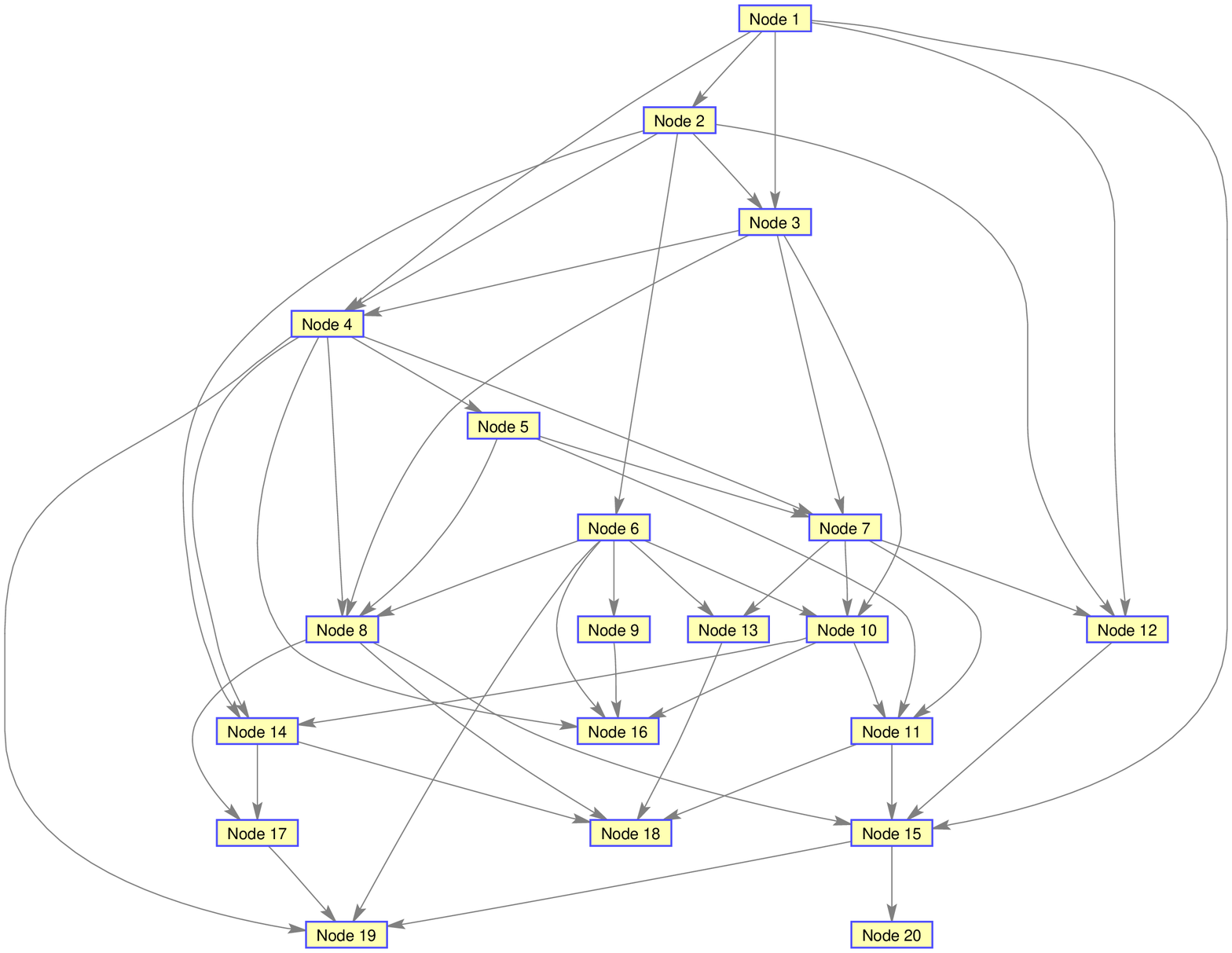}
	\caption{Example of random BAG with $20$ nodes and $m = 4$.}
	\label{exampleRandom}	
\end{figure}

To build this kind of BAGs we generate random DAGs where we limit the maximum number of parents that a node can have. This corresponds to restricting the maximum number of vulnerabilities that can lead an attacker to reach a certain security condition. Since in real scenarios we expect to have a reduced number of vulnerabilities that allows an attacker to compromise a network node \cite{whitehatsec_report,threat_report}, we consider that this structure is reasonable. Thus, for each node in the graph $X_i$, we randomly select its number of parents by drawing a random integer $n_p$ in the interval $[ 1, m ]$ uniformly, where $m$ is the maximum number of possible parents allowed. Then, we randomly select the $n_p$ parent nodes for $X_i$ from the set of nodes in the BAG for which $X_i$ is not a parent node already. This avoids directed cycles and preserves the DAG structure. In Fig.~\ref{exampleRandom} we show an example of a random BAG with $20$ nodes and $m = 4$. 

\subsubsection{Accuracy and Convergence}
In our first experiment we want to measure the accuracy and the convergence of both S-LBP and P-LBP for pseudo-random AGs. We have therefore generated synthetic AGs with $40$ nodes and $m=3$, where we have varied the proportion of OR and AND conditional probability tables. We have also explored different values for the damping factor $\alpha$ in the range $[0, 0.5]$. We have measured the accuracy using the Rooted Mean Squared Error (RMSE), comparing LBP estimates with the exact unconditional probabilities provided by JT. For each combination of parameters explored we have averaged the RMSE obtained for $20$ independent BAGs. From the results in Tab. \ref{tabAcc} we observe that the RMSE is less than $0.03$ in most cases, which is a reasonable accuracy to estimate the risks of compromising the different nodes in the AG, especially when considering that the probabilities of successful exploitation of vulnerabilities are not accurate, since they are estimated with the CVSS scores. Thus, the accuracy of LBP is enough to allow system administrators to decide the actions that need to be taken (both for the static and the dynamic analysis of the network). It is also interesting to observe that a little bit of damping, i.e. small values of $\alpha$, slightly improve the accuracy. Furthermore, it is remarkable that the RMSE is lower when the proportion of AND-type conditional probability tables is higher. This effect is due to the different coupling effect between the variables in the loops of the BAG depending on the type of conditional probability table. In Tab. \ref{tabAcc} we only show the RMSE for P-LBP, since the differences w.r.t. S-LBP were negligible. Moreover, we have also observed that both implementations of LBP converged in all cases.

\subsubsection{Accuracy with the Number of Iterations}
In our second experiment we have analysed how the accuracy of the LBP probability estimates changes with the number of iterations. LBP allows us to monitor the intermediate estimates of the unconditional probabilities, which is not possible with JT. This can help system administrators to reduce the time to respond to an attack, since they do not need to wait until the algorithm has completely converged. 

\begin{figure}
	\centering
	\psfrag{P-LBP}{{\scriptsize P-LBP}}
	\psfrag{S-LBP}{{\scriptsize S-LBP}}
	\psfrag{Iterations}[][]{{\scriptsize Iterations}}
	\psfrag{RMSE}[0.5cm][-0.5cm]{{\scriptsize RMSE}} 
	\includegraphics[width=7.5cm,height=4.7cm]{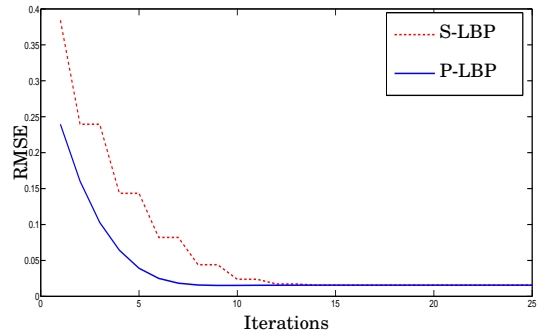}
	\caption{Average RMSE of P-LBP and S-LBP with the number of iterations for 25 pseudo-random BAGs with 100 nodes and $m=3$.}
	\label{resAccIter}	
\end{figure}

In Fig.~\ref{resAccIter} we show the average RMSE of P-LBP and S-LBP as a function of the number of iterations for 25 pseudo-random BAGs with 100 nodes and $m = 3$. The probability of having AND-type conditional probability tables has been set to $0.5$, and we have used a damping factor of $\alpha = 0.2$ in both LBP implementations. From the results in Fig.~\ref{resAccIter} we can observe that the algorithms converge on average in less than $15$ iterations, although P-LBP seems to converge faster, and gets better estimates of the unconditional probabilities after the first iteration (although the final result is similar to S-LBP). From Fig.~\ref{resAccIter} it is important to note that, after 5 iterations, the RMSE for P-LBP is about $0.05$. This accuracy can be considered reasonable to start planning risk mitigation strategies at run-time without waiting for LBP to converge.

\subsubsection{Time Scalability}
Our last experiment with pseudo-random BAGs is aimed at evaluating the time scalability of LBP and JT for the static and dynamic analysis of AGs. We have analysed networks with different sizes and different densities: For $m$ we have explored the values $3$ and $4$, while for $n$, the number of nodes in the BAG, we have used values in the range $[20, 3000]$. However, for JT we have limited the value of $n$ to $120$ for $m = 3$ and to $80$ for $m = 4$ because of physical memory limitations\footnote{The experiments have been conducted in a 16 GB computer with an Intel Core i7 processor at 3.40 GHz.}. The probability of having AND-type conditional probability tables is set to $0.5$. For each value of $n$ and $m$ we have generated $20$ pseudo-random BAGs and, for each BAG, we compute the unconditional probabilities for all the nodes with both LBP and JT.

\begin{figure}
	\centering
	\psfrag{P-LBP-3}{{\scriptsize P-LBP-3}}
	\psfrag{P-LBP-4}{{\scriptsize P-LBP-4}}
	\psfrag{S-LBP-3}{{\scriptsize S-LBP-3}}
	\psfrag{S-LBP-4}{{\scriptsize S-LBP-4}}
	\psfrag{JT-3}{{\scriptsize JT-3}}
	\psfrag{JT-4}{{\scriptsize JT-4}}
	\psfrag{Nodes}[][]{{\scriptsize Number of nodes}}
	\psfrag{Time}[0.5cm][-0.5cm]{{\scriptsize Time (s)}} 
	\includegraphics[width=8.5cm,height=5.5cm]{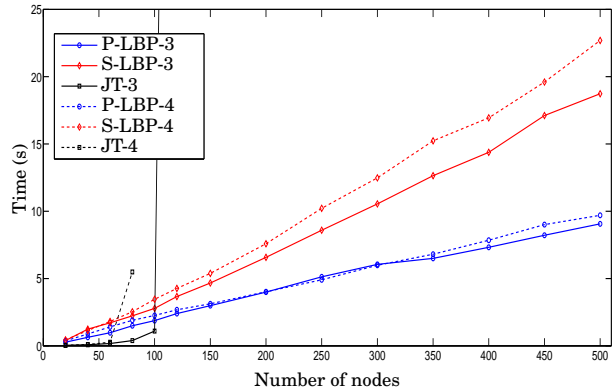}
	\caption{Average time to compute the unconditional probabilities for P-LBP, S-LBP, and JT for pseudo-random attack graphs. The notation P-LBP-$m$, S-LBP-$m$, and JT-$m$ stands for the value of $m$, the maximum number of parents allowed for each node, used to generate the graphs in each case.}
	\label{resFig1}	
\end{figure}

In Fig.~\ref{resFig1} we show the average time required to compute the unconditional probabilities for all the nodes in the BAG for P-LBP, S-LBP, and JT. In the case of JT, the measured time includes the time required to build the clique tree, and compute all the messages and all the unconditional probabilities. For LBP, this time considers the computation of all the messages and the probability estimates. Therefore, we are, in essence, measuring the time required to perform the static analysis of the BAG. For both LBP variants we use $\alpha = 0.2$ and set the maximum number of iterations to two times the number of nodes in the network. As in the previous experiment, both LBP implementations converged in all cases.

From the results shown in Fig.~\ref{resFig1} we can observe the exponential increase of JT with the number of nodes, whereas both P-LBP and S-LBP scale linearly. Although the time to compute the unconditional probabilities by JT is lower for small AGs (less than 100 nodes), it appears that LBP is a suitable alternative to make inference in large BAGs, where the exponential scalability of JT makes its use impractical. It is also interesting to note that P-LBP is faster than S-LBP. Moreover, when increasing the complexity of the network (by increasing $m$) the performance of P-LBP remains similar, whereas we can observe larger differences for S-LBP. 

For the dynamic analysis of AGs, when we observe evidence of compromise in some nodes, we need to recompute all the messages taking into account the evidence as well as the posterior probabilities in all the graph nodes. This applies equally well to JT and LBP. However, for JT, we do not need to build the clique tree again. So, to analyse the performance of the algorithms for the dynamic analysis of AGs, we have randomly selected $3$ nodes in each graph where we consider that evidence of attack has been observed. Then, we have measured the time required to recompute the posterior probabilities for P-LBP, S-LBP, and JT. We have omitted the resulting figure, since the results are very similar to those obtained for the static analysis (shown in Fig.~\ref{resFig1}). The measured times for the dynamic analysis are slightly lower for all the methods and the differences are not very significant. In the case of the JT, this means that for this kind of graphs the bottleneck of the algorithm is the computation of the messages rather than the time required to build the clique tree \cite{luis}. This is due to the strong interconnection of the nodes in the graph, which makes some clusters in the clique tree to have a high number of variables. For LBP implementations, the similarity of the results suggests that the number of iterations needed to converge are similar.

For the sake of clarity, we have omitted in Fig.~\ref{resFig1} the results obtained with LBP for networks from $500$ to $3000$ nodes. However, we have observed that the linear scalability of LBP holds. Thus, for BAGs with $3000$ nodes the average time to perform the static (or dynamic) analysis or the graph is about $60$ seconds on a standard laptop. The linear scalability of LBP for this kind of graphs make it useful for both static and dynamic analysis of AGs, especially when the graphs are large. It is thus possible to analyse AGs with thousands of nodes, which can correspond to networks with tens or hundreds of thousands of nodes (depending on the number of vulnerabilities).

\subsection{BAGs with cluster structure}
\begin{figure}
	\centering
	\includegraphics[width=12cm,height=7.5cm]{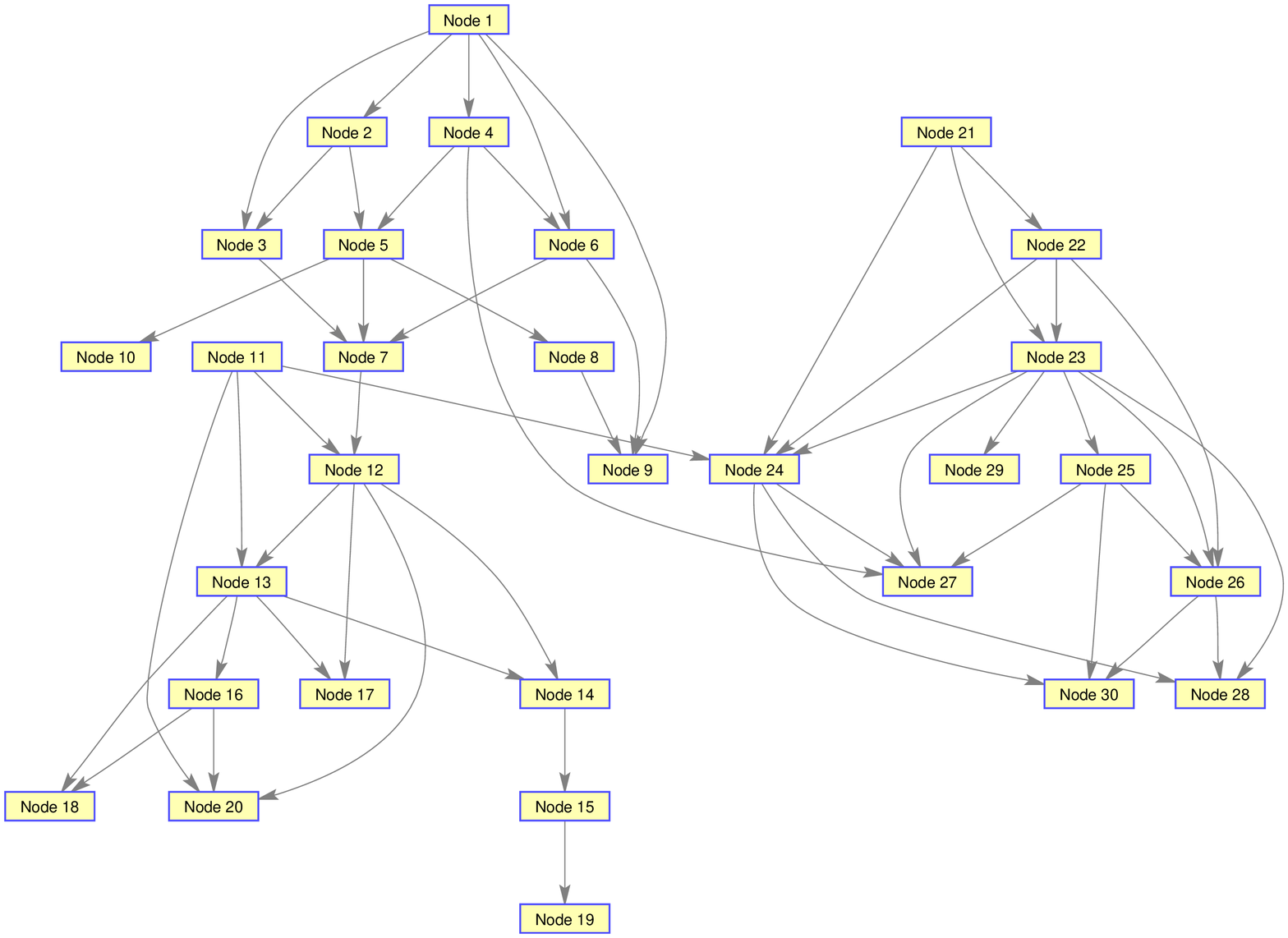}
	\caption{Example of clustered BAG with 3 subnetworks and 10 nodes per subnetwork.}
	\label{exampleCluster}	
\end{figure}

For the experimental evaluation we have also studied the effect of clustering on the analysis by using synthetic graphs with a cluster structure. Typical corporate networks are structured into subnetworks \cite{Tan2003} and contain several hosts with common software installations so we can expect some form of cluster structure in the corresponding AG. To generate this kind of graphs we have considered networks with clusters of the same size $n_c$. Then, for each cluster, following the same procedure as before, we have generated pseudo-random subgraphs, limiting the maximum number of parents for each node to $m$. Finally, to include the dependencies between clusters, we have added one edge from one node in each cluster to one node in each of the other clusters, provided that the DAG structure required for BNs is preserved. For our experiments we have generated synthetic clustered graphs with $n_c = 20$ and $50$, varying the total number of network nodes from $100$ to $1000$. In Fig.~\ref{exampleCluster} we show an example of a clustered BAG with $3$ subnetworks, with $10$ nodes per subnetwork, and $m = 3$. 

For this kind of BAGs we have performed the same experiments as for pseudo-random BAGs. Thus, we first assess the accuracy and convergence of the two LBP implementation; second, we show how the accuracy evolves with the number of iterations and, finally, we show the time scalability for the static and dynamic analysis of the graphs.

\subsubsection{Accuracy and Convergence}
For this experiment we have generated synthetic cluster AGs with $5$ clusters with $n_c = 20$ nodes per cluster and $m=3$. As for the pseudo-random BAGs, we have varied the proportion of OR and AND conditional probability tables and we have explored values for the damping factor $\alpha$ in the range $[0, 0.5]$. For each combination of parameters explored we have averaged the RMSE obtained for $20$ independent BAGs. The results are shown in Tab.~\ref{tabAccCluster}, where we observe that the average RMSE is less than $0.04$ in all cases, which means that the probability estimates provided by LBP are reasonable to perform risk assessment with BAGs. We can also appreciate that the proportion of AND/OR conditional probability tables has some impact on the accuracy of the probability estimates. Thus, having a bigger proportion of AND-type conditional probability tables results in more accurate estimations for the unconditional probabilities in the BAG. In this case, in contrast to the results shown in Tab.~\ref{tabAcc}, damping does not provide any clear improvement in the accuracy. However, even in these cases, it is a good practice to include a little bit of damping in LBP updates to avoid potential instabilities in the algorithm. As in the case of the pseudo-random BAGs, we only show the RMSE for P-LBP, since the results obtained from S-LBP are very similar. Finally, we have also observed that both, P-LBP and S-LBP, converged in all cases.

\begin{table}
\tbl{Average RMSE plus/minus one standard deviation for P\-LBP on clustered BAGs with $m=3$, $5$ subnetworks and $n_c = 20$ nodes per subnetwork, varying the damping factor $\alpha$ and the probability of having AND-type conditional probability tables, p(AND). \label{tabAccCluster}}{
\begin{tabular}{|c|c|c|c|c|c|}
\hline
p(AND) & $\alpha = 0.0$ & $\alpha = 0.1$ & $\alpha = 0.3$ & $\alpha = 0.5$ \\
\hline
\hline
$0.0$ & $0.0342 \pm 0.0213$ & $0.0376 \pm 0.0213$ & $0.0341 \pm 0.0157$ & $0.0337 \pm 0.0186$ \\
\hline
$0.2$ & $0.0216 \pm 0.0078$ & $0.0247 \pm 0.0105$ & $0.0289 \pm 0.0154$ & $0.0229 \pm 0.0123$ \\
\hline
$0.5$ & $0.0185 \pm 0.0062$ & $0.0219 \pm 0.0128$ & $0.0193 \pm 0.0083$ & $0.0224 \pm 0.0091$ \\
\hline
$0.8$ & $0.0149 \pm 0.0042$ & $0.0136 \pm 0.0035$ & $0.0138 \pm 0.0078$ & $0.0127 \pm 0.0053$ \\
\hline
$1.0$ & $0.0109 \pm 0.0042$ & $0.0115 \pm 0.0039$ & $0.0093 \pm 0.0043$ & $0.0107 \pm 0.0038$ \\
\hline
\end{tabular}}
\end{table}

\subsubsection{Accuracy with the Number of Iterations}
\begin{figure}
	\centering
	\psfrag{P-LBP}{{\scriptsize P-LBP}}
	\psfrag{S-LBP}{{\scriptsize S-LBP}}
	\psfrag{Iterations}[][]{{\scriptsize Iterations}}
	\psfrag{RMSE}[0.5cm][-0.5cm]{{\scriptsize RMSE}} 
	\includegraphics[width=7.5cm,height=4.7cm]{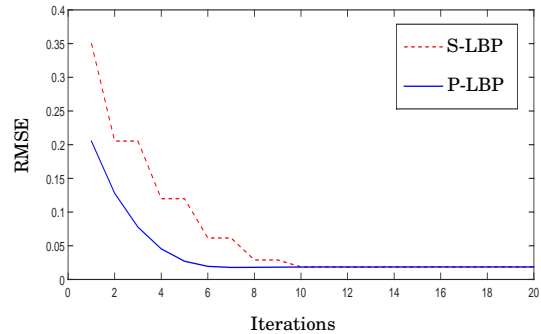}
	\caption{Average RMSE of P-LBP and S-LBP with the number of iterations for 25 cluster BAGs with 100 nodes (5 subnetworks with 20 nodes per subnetwork) and $m=3$.}
	\label{resAccIterCluster}	
\end{figure}

In Fig.~\ref{resAccIterCluster} we show how the average RMSE of P-LBP and S-LBP decreases with the number of iterations. For these experiments we have generated 25 cluster BAGs with $5$ clusters and $n_c = 20$ nodes per cluster, with $m = 3$. As in the case of the pseudo-random BAGs, we have set the probability of having AND-type conditional probability tables to $0.5$ and we have used a damping factor of $\alpha = 0.2$. 

The results in Fig.~\ref{resAccIterCluster} are very similar to those shown in Fig.~\ref{resAccIter} for the case of the pseudo-random BAGs. Thus, we can observe that after $4$ iterations, P-LBP produce estimates for the unconditional probabilities with an average RMSE less than $0.05$. We also appreciate that, as in the previous case, P-LBP converges faster than S-LBP, although both techniques achieve a similar approximation error after convergence. The result of this experiment shows that, after a few iterations, the accuracy provided by LBP can be considered reasonable to start planning risk mitigation strategies at run-time before LBP converges.

\subsubsection{Time Scalability}
\begin{figure}
	\centering
	\psfrag{P-LBP-20}{{\scriptsize P-LBP-20}}
	\psfrag{P-LBP-50}{{\scriptsize P-LBP-50}}
	\psfrag{S-LBP-20}{{\scriptsize S-LBP-20}}
	\psfrag{S-LBP-50}{{\scriptsize S-LBP-50}}
	\psfrag{JT-20}{{\scriptsize JT-20}}
	\psfrag{JT-50}{{\scriptsize JT-50}}
	\psfrag{Nodes}[][]{{\scriptsize Number of nodes}}
	\psfrag{Time}[0.5cm][-0.5cm]{{\scriptsize Time (s)}} 
	\includegraphics[width=8cm,height=5cm]{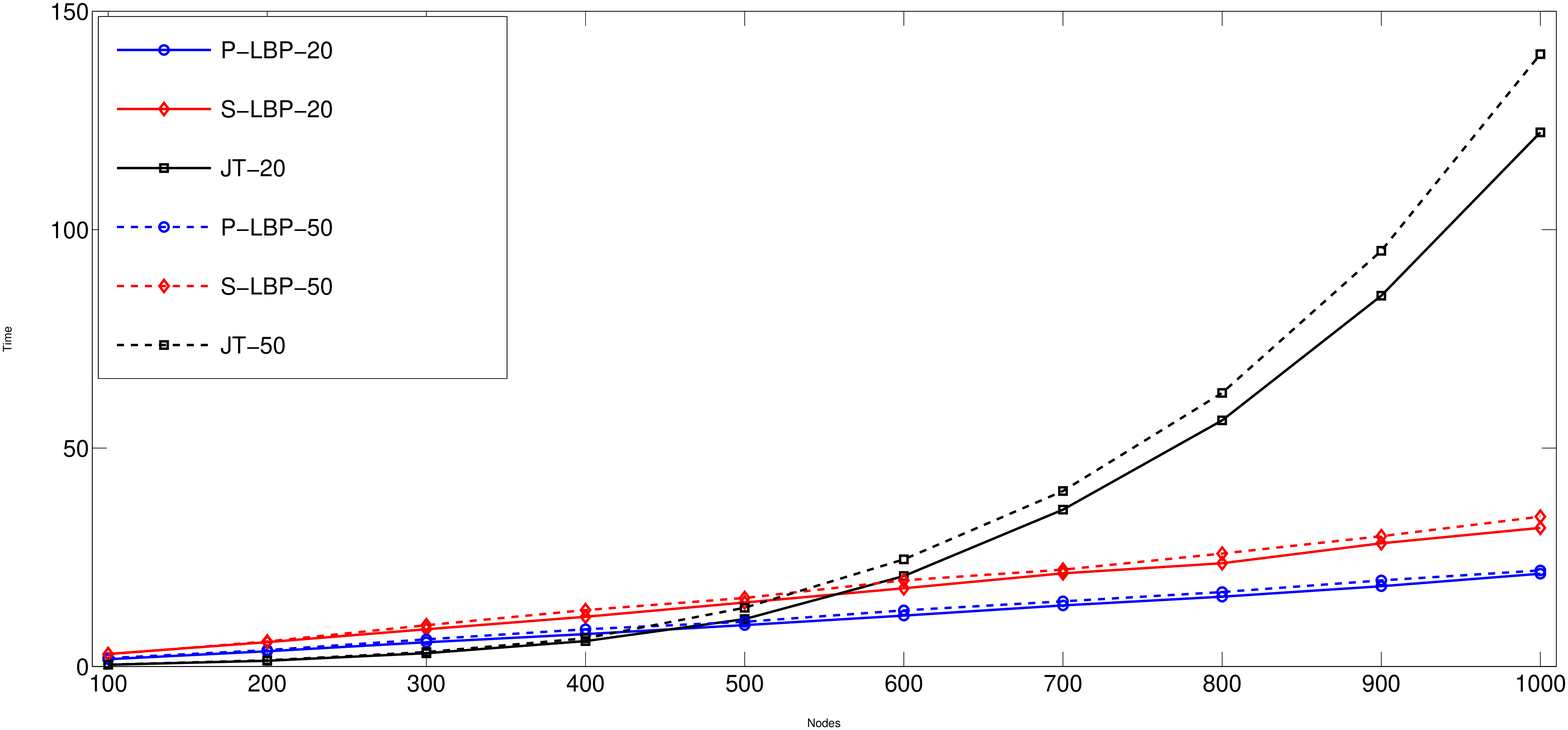} \\
	(a) \vspace{0.3cm} \\
	\includegraphics[width=8cm,height=5cm]{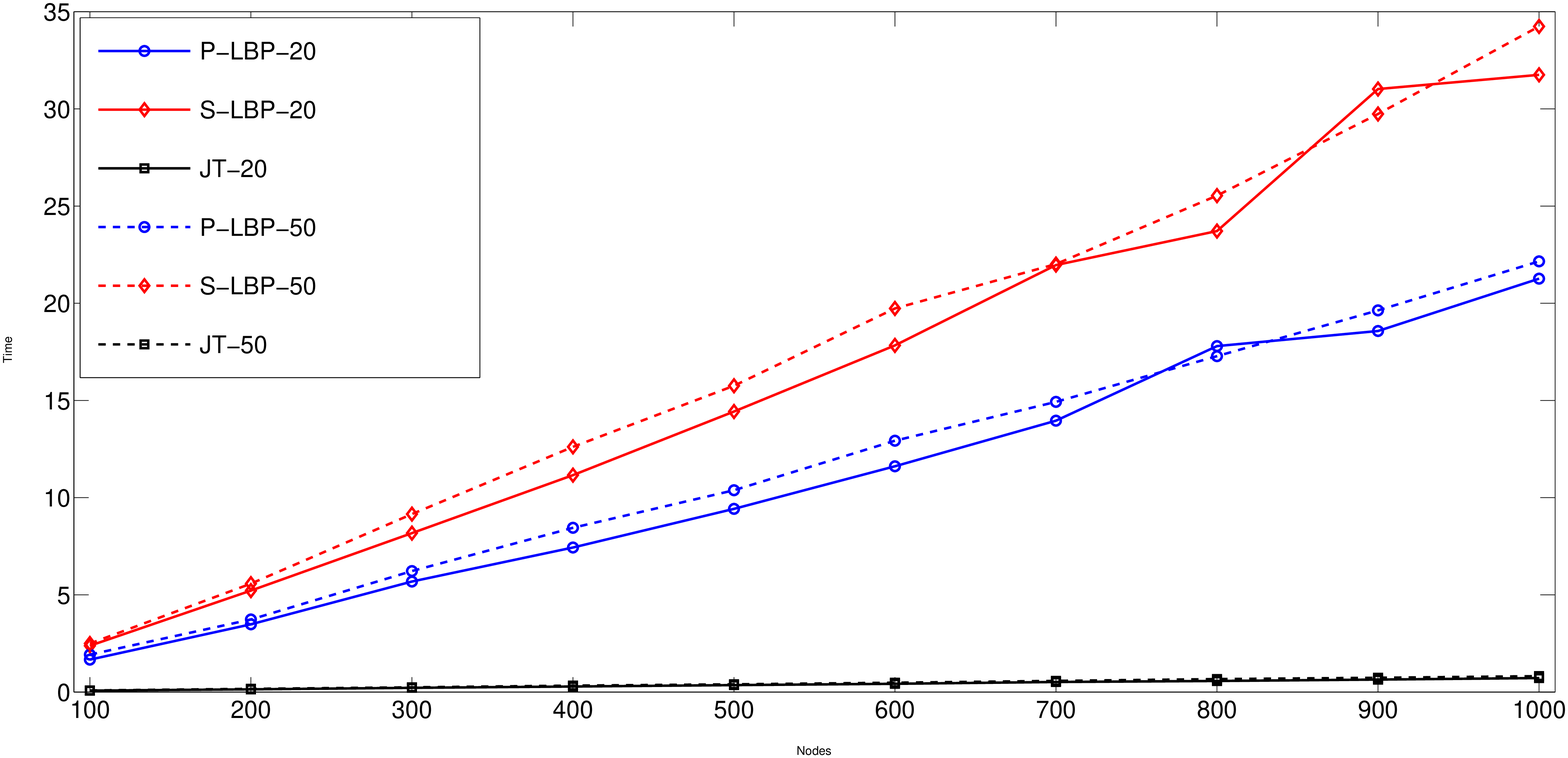} \\
	(b)  \\
	\caption{Time to compute the unconditional probabilities for P-LBP, S-LBP, and the JT algorithm for cluster networks with different cluster sizes ($n_c = 20$ and $50$) and $m = 3$ for: (a) the static analysis and (b) the dynamic analysis (when we observe evidence of attack at $3$ random nodes).}
	\label{resFig2}	
\end{figure}

We report in Fig.~\ref{resFig2}.(a) the time required to perform the static analysis for P-LBP, S-LBP, and JT. As in the case of the pseudo-random networks we observe that JT scales exponentially with the number of nodes, although it is able to perform static analysis on larger networks (compared to the pseudo-random AGs). In contrast, both LBP implementations scale linearly with the number of nodes in the BAG, and require less time than JT to compute all the unconditional probabilities for graphs with more than $500$ nodes. Also in line with the previous experiment, P-LBP shows a better performance than S-LBP, although the difference is not as significant as before. We can also appreciate that, for both LBP methods, the increment on the size of the clusters implies only a small difference in the time performance of the algorithms. 

In Fig.~\ref{resFig2}.(b) we show the time required to perform the dynamic analysis when we observe evidence of attack in $3$ nodes chosen at random. We can observe again that the scalability for both LBP implementations is linear in the number of nodes, and that the time required to compute all the unconditional probabilities in the BAG is slightly lower than in the case of the static analysis in Fig.~\ref{resFig2}.(a). However, in this case the performance of JT is also linear in the number of nodes and the time required to compute the unconditional probabilities is much lower than in the case of two LBP methods. 

These results indicate that, when JT is applied on clustered networks, the bottleneck is the generation of the clique tree, while the computation of the messages is simple. This suggests that the cluster structure of the BAGs produces clique trees with a reduced number of variables in each cluster of the tree. Hence, the messages sent across nodes involve a reduced number of random variables, allowing a fast calculation of the posterior probabilities for the dynamic analysis of the BAGs. In the case of LBP, the messages are simpler than in the case of JT. However, the time to compute the unconditional probabilities in the dynamic analysis is higher since, at each iteration, we need to compute all the messages and several iterations are needed to make the algorithm converge. Although JT is faster than LBP for the dynamic analysis of clustered BAGs, the exponential scalability for the static analysis, when the clique tree is generated, limits the tractability of the analysis for large networks. On the other hand, as mentioned before, LBP allows us to monitor the values of the posterior probabilities at each iteration, so that we can obtain accurate estimates for the probabilities before the algorithm converges. 


\section{Conclusions}
Bayesian Networks are a powerful tool for the static and dynamic analysis of AGs: the unconditional probabilities at each node in the graph (not only for the target) provide useful information to system administrators for security risk assessment and mitigation. The values of these probabilities take into account the dependencies between the different attack paths and the difficulty of exploiting each vulnerability, in contrast to other measures proposed in the literature to perform risk assessment in AGs. However, the scalability problems of the exact inference techniques proposed in the literature for the static and dynamic analysis of BAGs limit the applicability of these techniques to graphs with a few hundreds of nodes, far from the size of the AGs we can expect in large corporate networks.

In this paper we have shown that LBP, an approximate inference technique, can be used effectively for the static and dynamic analysis of large BAGs. We have verified through experiments that the reduction in the computational cost and memory requirements is significant. Moreover our experimental evaluation shows that \emph{LBP scales linearly with the number of nodes for both the static and the dynamic analysis}. Overall these results show significantly better performance than the techniques proposed for the analysis of BAGs in the literature so far. We have experimented with synthetic AGs with a broad variety of topologies in an effort to ensure the applicability of the technique to many network deployments. The gains in scalability are obtained at the price of a loss in accuracy on the probabilities. We have however verified through experiments that this accuracy loss is very low with an average RMSE of less than $0.03$, especially taking into account that the values for the probability of successful exploitation of the vulnerabilities, used to build the BAG models, are non-accurate estimates. We have also shown through the experiments that LBP compares favourably with the JT, the state of the art technique for exact inference in BAGs \cite{luis}. Although a significant amount of literature on the application of AG exists, few studies have considered the computational aspect of making inference on them. Furthermore, the lack of scalability has significantly hindered their application. Our results show that by using the right techniques, both static an dynamic analysis can be performed on AGs with thousands of nodes, even on a standard laptop computer. We have also evaluated the effect of clustering on the performance of the analysis and shown that this can lead to further significant gains in performance. Our future research plans include modelling the attacker's capabilities to estimate the probability of successful exploitation of vulnerabilities, and the use of Bayesian inference techniques to help prioritizing forensic investigation using AGs.


%

\begin{acks}
The authors would like to thank British Telecom for their collaboration in this research and our colleagues in our research group for their contribution to this work through many useful discussions. This research has been funded by the UK government under EPSRC project EP/L022729/1.
\end{acks}

\bibliographystyle{ACM-Reference-Format-Journals}
\bibliography{biblio}

\received{June 2016}{June 2016}{June 2016}
%
%
%
%
%
%
%

\end{document}